\RequirePackage{fix-cm}
\documentclass[twocolumn]{svjour3}          
\smartqed  
\usepackage{graphicx}

\usepackage{listings}
\usepackage{xcolor}
\usepackage{textcomp}
\usepackage{todonotes}
\usepackage{url}

\usepackage[utf8]{inputenc}

\usepackage{relsize}

\usepackage{framed}
\usepackage{color}

\usepackage{enumerate}

\usepackage{placeins}
\usepackage{xspace}

\usepackage{tabularx}

{\begin{snugshade}\begin{quote}\vspace{1.2em}}
{\hfill\vspace{1.2em}\end{quote}\end{snugshade}}

\definecolor{shadecolor}{rgb}{0.9,0.9,0.9}

\journalname{Technical Report}
\begin{document}

\title{A Generative Approach for User-Centered, Collaborative, Domain-Specific Modeling Environments}



\author{Philip Zweihoff\thanks{\emph{Correspondence to:} philip.zweihoff@tu-dortmund.de}, %
Bernhard Steffen
}

\institute{Chair for Programming Systems, Department of Computer Science, TU Dortmund University, 44227 Dortmund, Germany}



\date{6th April 2021}

\authorrunning{P. Zweihoff et al.}

\maketitle

\definecolor{numbergray}{gray}{0.5}

\lstdefinestyle{mgl}{
   columns=fixed,
	breaklines=true,
	numbers=left,
	numberstyle=\tiny,
	stepnumber=1,
	numbersep=5pt,
   showspaces=false,
   showstringspaces=false,
   frame=tblr,
   tabsize=2,
   frame=shadowbox,
   columns=fixed,
   basicstyle=\ttfamily\scriptsize,
   rulesepcolor=\color[rgb]{0.6, 0.6, 0.6},
    keywordstyle=\color[rgb]{0.5,0,0.34}\textbf,
   stringstyle=\color{blue},
   commentstyle=\color[rgb]{0.4,0.7,0.4},
   backgroundcolor=\color[rgb]{0.97,0.97,0.97},
   morekeywords={node, edge, attr, container, as, package, nsURI, iconPath,
	diagramExtension, graphModel, incomingEdges, outgoingEdges, for, prime,
	parameters, names, types, extends, abstract},
	morestring=[b]",
   showtabs=false,
	morecomment=[l]{//},
	morecomment=[s]{/*}{*/},
   literate={0}{{{\color{numbergray}0}}}{1}%
		{1}{{{\color{numbergray}1}}}{1}%
		{2}{{{\color{numbergray}2}}}{1}%
		{3}{{{\color{numbergray}3}}}{1}%
		{4}{{{\color{numbergray}4}}}{1}%
		{5}{{{\color{numbergray}5}}}{1}%
		{6}{{{\color{numbergray}6}}}{1}%
		{7}{{{\color{numbergray}7}}}{1}%
		{8}{{{\color{numbergray}8}}}{1}%
		{9}{{{\color{numbergray}9}}}{1}
}

\newcommand{\includemgl}[3]{\lstinputlisting[style=mgl, float=b, caption=#2, label=#3, captionpos=b]{#1}}

\lstdefinestyle{style}{
   columns=fixed,
	breaklines=true,
	numbers=left,
	numberstyle=\tiny,
	stepnumber=1,
	numbersep=5pt,
   showspaces=false,
   showstringspaces=false,
   frame=tblr,
   tabsize=2,
   frame=shadowbox,
   columns=fixed,
   basicstyle=\ttfamily\scriptsize,
   rulesepcolor=\color[rgb]{0.6, 0.6, 0.6},
   keywordstyle=\color[rgb]{0.5,0,0.4}\textbf, 
   stringstyle=\color{blue},
   commentstyle=\color[rgb]{0.4,0.7,0.4},
   backgroundcolor=\color[rgb]{0.97,0.97,0.97},
	morekeywords={nodeStyle, edgeStyle, rectangle, ellipse, roundedRectangle,
	text, polyline, size, corner, position, value, color, lineStyle, lineWidth,
	SOLID, DASH, DASHDOT, DASHDOTDOT, DOT, decorator, location, relativeToMid,
	points, appearance, appearanceProvider, background, foreground, relativeTo,
	ARROW, CIRCLE, TRIANGLE, CENTER, MIDDLE, @, extends, movable},
	morestring=[b]",
   showtabs=false,
	morecomment=[l]{//},
	morecomment=[s]{/*}{*/},
   literate={0}{{{\color{numbergray}0}}}{1}%
		{1}{{{\color{numbergray}1}}}{1}%
		{2}{{{\color{numbergray}2}}}{1}%
		{3}{{{\color{numbergray}3}}}{1}%
		{4}{{{\color{numbergray}4}}}{1}%
		{5}{{{\color{numbergray}5}}}{1}%
		{6}{{{\color{numbergray}6}}}{1}%
		{7}{{{\color{numbergray}7}}}{1}%
		{8}{{{\color{numbergray}8}}}{1}%
		{9}{{{\color{numbergray}9}}}{1}
}

\newcommand{\includestyle}[3]{\lstinputlisting[style=style, float=b, caption=#2, label=#3, captionpos=b]{#1}}

\lstdefinestyle{java}{
   language=Java,
   breaklines=true,
   numbers=left,
   numberstyle=\tiny,
   stepnumber=1,
   numbersep=5pt,
   showspaces=false,
   showstringspaces=false,
   frame=tblr,
   tabsize=2,
   frame=shadowbox,
   columns=fixed,
   basicstyle=\ttfamily\scriptsize,
   rulesepcolor=\color[rgb]{0.6, 0.6, 0.6},
   keywordstyle=\color[rgb]{0.5,0,0.4},
   stringstyle=\color{blue},
   commentstyle=\color[rgb]{0.4,0.7,0.4},
   backgroundcolor=\color[rgb]{0.97,0.97,0.97},
   showtabs=false
}

\newcommand{\includejava}[3]{\lstinputlisting[style=java, float=b, caption=#2, label=#3, captionpos=b]{#1}}

\lstdefinestyle{xtext}{
   columns=fixed,
        breaklines=true,
        numbers=left,
        numberstyle=\tiny,
        stepnumber=1,
        numbersep=5pt,
   showspaces=false,
   showstringspaces=false,
   frame=tblr,
   tabsize=2,
   frame=shadowbox,
   columns=fixed,
   basicstyle=\ttfamily\scriptsize,
   rulesepcolor=\color[rgb]{0.6, 0.6, 0.6},
   keywordstyle=\color[rgb]{0.5,0,0.34}\textbf,
   stringstyle=\color{blue},
   commentstyle=\color[rgb]{0.4,0.7,0.4},
   backgroundcolor=\color[rgb]{0.97,0.97,0.97},
   morekeywords={grammar, with, hidden, generate, as, import, returns, current, terminal, enum, STRING, ID},
   keywordstyle=[2]{\textbf},
   morecomment=[l]{//}, 
   morecomment=[s]{/*}{*/}, 
   morestring=[b]",
   morestring=[b]',
   tabsize=2
}

\newcommand{\includextext}[3]{\lstinputlistinforEachg[style=xtext, float=tb, caption=#2, label=#3, captionpos=b]{#1}}

\definecolor{extension}{RGB}{174,48,0}

\lstdefinestyle{xtend}{
	language=Java,
	breakatwhitespace=true,
   columns=fixed,
        breaklines=true,
        numbers=left,
        numberstyle=\tiny,
        stepnumber=1,
        numbersep=5pt,
   showspaces=false,
   showstringspaces=false,
   frame=tblr,
   tabsize=2,
   frame=shadowbox,
   columns=fixed,
   basicstyle=\ttfamily\scriptsize,
   rulesepcolor=\color[rgb]{0.6, 0.6, 0.6},
   keywordstyle=\color[rgb]{0.5,0,0.34}\textbf,
   stringstyle=\color{blue},
   commentstyle=\color[rgb]{0.4,0.7,0.4},
   backgroundcolor=\color[rgb]{0.97,0.97,0.97},
 morekeywords={override,def,val,cached,case,default,extension,false,import,JAVA,WORKFLOWSLOT,let,new,null,private,create,switch,this,true,reexport,around,if,then,else,context},
 keywordstyle=[2]{\textbf},
 morecomment=[l]{//}, 
 morecomment=[s]{/*}{*/}, 
 morestring=[b]",
 tabsize=2,
 literate={filter}{{\textcolor{extension}{filter}}}{5}
          {empty}{{\textcolor{extension}{empty}}}{4}
}

\newcommand{\includextend}[3]{\lstinputlisting[style=xtend, float=tb, caption=#2, label=#3, captionpos=b]{#1}}

\hyphenation{Fow-ler Bar-ce-lo-na Pou-na-mu Pro-ject ap-proa-ches Archi-medean with-in}

\begin{abstract}
The use of low- and no-code modeling tools is today an established way in practice to give non-programmers an opportunity to master their digital challenges independently, using the means of model-driven software development.
However, the existing tools are limited to a very small number of different domains such as mobile app development, which can be attributed to the enormous demands that a user has on such a tool today.
These demands exceed the mere use of a modeling environment as such and require cross-cutting concerns such as: easy access, direct usability and simultaneous collaboration, which result in additional effort in the realization of such tools.

Our solution is based on the idea to support and simplify the creation of new domain-specific holistic tools by generating it entirely based on a declarative specification with a domain-specific meta-tool.
The meta-tool Pyro demonstrated and analyzed here focuses on graph-based graphical languages to fully generate a complete, directly executable tool starting from a metamodel in order to meet all cross-cutting requirements.
\end{abstract}


\section{Introduction}
In the context of digitalization, the need for and importance of individual software solutions is constantly growing.
Be it to handle internal processes more efficiently, to integrate new digital solutions into an existing system or to expand into new business areas.
The associated dependence on programmers, service providers or software manufacturers presents many companies with major problems, so that almost all corporations are forced to deal with software development themselves.

The only options are to acquire an existing solution and to customize it to one's own needs or to carry out a development from scratch in order to cover special requirements.
The biggest hurdle here is the requirements analysis, which represents the translation of the goals to be achieved into a concrete technical implementation.
In order to perform a requirements analysis successfully, it is necessary that both parties speak the same language in order to leave no room for misunderstandings and misinterpretations.
In practice, however, this language is rarely given, since a requirement rather formulates "what" should work in the end and less "how" it is concretely implemented. 
These different languages result from the technical background of the parties involved, in that on the one hand there is exclusively domain knowledge and on the other hand a technically and generically trained software developer.
This results in the so-called "semantic gap", \cite{margaria2007service} which must be overcome again for each requirements analysis and review.

Low- and no-code tools, which allow domain experts without programming knowledge to independently implement solutions for their separate problems, are now a widespread alternative.
These tools rely on methods of the model-driven software development (MDSD) \cite{stahl2006model} and offer domain-specific languages (DSLs) \cite{fowler2010domain} for the definition of a solution.
Extensions of the concept of MDSD like the eXtreme model-driven development (XMDD) \cite{margaria2020extreme,margaria2012service} have already shown in several studies by Lamprecht, Margarian and Saay \cite{9156062,lamprecht2014scientific,lamprecht2015use} that the use of domain-specific languages is an effective means to give non-programmers an intuitive approach to technical problems. 
A DSL uses constructs, notations, and associated specialization specific to a domain to best support the user.
DSLs can have both textual and graphical syntax, depending on the domain and the notations known there.
However, several studies \cite{chen2019effects,horn2009comparing,moher1993comparing} already showed that graphical DSLs are easier to learn and more intuitive to use.
To finally reach a technical implementation, machine translators in the form of generators, interpreters and transformers are used, which create the bridge between the domain expert and an actual realization.
As a result, users can focus on the "what," with the "how" anchored in the particular tool.
This concept of language-driven-engineering (LDE) \cite{steffen2019language} and the related model-driven engineering (MDE)  \cite{schmidt2006model} has already been successfully applied in various courses \cite{gossen2018dsls,margaria2018generative,tegeler2019product} and projects \cite{howar2019jconstraints,gossen2019large}. 

Conversely, however, this approach requires that an individual tool be available for each individual domain so that the respective notation and semantics are available.
The development and provision of such domain-specific tools is usually complex and time-consuming.
In addition to the design of the syntax and semantics of a domain-specific language, many technical requirements must be met in order to give users the most straightforward and intuitive access possible to the new tool.

The requirements of simple and direct accessibility result in the need for a distributed and browser-accessible application as the basis of the tool.
The tool should not need to be installed and set up, and should not impose any separate requirements on the user's system.
In addition, users today are accustomed to simultaneous collaboration without further intervention. 
Accordingly, for the development of any concrete tool, the various technical problems have to be solved again and again.

In contrast, Pyro \cite{zweihoff2019pyro,lybecait2018tutorial} represents an archimedean point in the sense of \cite{steffen2016archimedean} by supporting and simplifying the entire development process of DSLs.
Pyro focuses on the implementation and deployment of graph-based DSLs for use in a web context.
For the development of a new DSL, only two Ecore-based metamodels \cite{gronback2009eclipse} are initially required for the declaration of the concrete and abstract syntax.
Since Pyro is limited to graph-based graphical languages, the metamodel of a DSL extends the meta-types Graphmodel, Node, Container and Edge.
The creation of Ecore metamodels in the context of the Eclipse Modeling Framework (EMF) \cite{steinberg2008emf} is supported by various tools such as Xtext \cite{bettini2016implementing}, and Emoflon \cite{anjorin2011emoflon}, which differ in their ease of entry and intuitiveness.
In this paper, the realization of the metamodel with the CINCO Meta Tooling Suite \cite{naujokat2018cinco} is demonstrated.
CINCO is characterized by a simplified declarative creation of metamodels by using its own DSL.

Starting from the metamodel of the DSL, Pyro creates a complete, directly usable, web-based modeling platform in one generation step, so that users can directly access the newly created languages via their browser without any installation.
The underlying generated distributed system enables simultaneous collaboration between users and provides a customized editor for modeling in the respective DSL.

Starting with the listing of the various current requirements for a domain-specific tool in section \ref{sec:user_requirements} and \ref{sec:development_challenges}, the following sections explain which approaches and technical concepts are implemented in Pyro to meet the various challenges.
For this purpose, section \ref{sec:dsl_tool_development} first explains how a new DSL can be specified.
Subsequently, section \ref{sec:dsl_tool_development} describes how the generated web-based modeling environment is implemented and which technical concepts have been realized.
A special challenge is the simultaneous collaboration, the realization of which will be discussed in section \ref{sec:simultaneous_collaboration}.
In order to highlight the special features of Pyro, a comparison with similar tools in section \ref{sec:tool_comparison} is performed.
Finally, the concepts and technical aspects of Pyro are summarized and an outlook on future developments is given in section \ref{sec:summary_and_outlook}.

\section{User Requirements}
\label{sec:user_requirements}
Regardless of whether a modeling environment or an application is being developed, user acceptance is the decisive factor for success \cite{dasgupta2007user,venkatesh2003user}.
In order to actually support the user and generate added value, features on the one hand and a customized user interface on the other must be realized.

This is especially true for domain-specific tools, which are developed along the concept of Fowler \cite{fowler2010domain} explicitly for a specific target group.
In addition to the DSL as the core of a tool, additional requirements may arise for the system itself, depending on the user group.
These requirements in turn give rise to a variety of technical challenges that would have to be overcome in the realization of a new tool.

Currently, DSLs are largely provided and used with the help of established IDEs such as Eclipse \cite{burnette2005eclipse}, Theia \cite{domros2018moving}, Visual Studio Code \cite{wwwvscode} and IDEA MPS \cite{pech2013jetbrains} from the programming environment.
These enriched general purpose IDEs provide versatile support and have high acceptance among engineers of various disciplines.
However, these environments appear overwhelming and cluttered to users from non-engineering domains, which are, however, the ideal target group for the use of DSLs, as already shown by X \cite{nguyen2014learning,otaduy2017user,wile2004lessons}.
The analysis presented in this chapter attempts to formulate the current requirements for a DSL tool that addresses the needs of non-technical users:

\subsubsection*{Graphical Language}

The use of domain-specific languages requires the emulation or adaptation of known and established notation and representation forms from the domain.
Both in engineering and in the environment of non-technical users, graphical languages such as UML \cite{bruegge2009object}, BPMN \cite{white2004introduction}, and AADL \cite{feiler2006architecture} are predominantly used to describe and illustrate complex relationships.
For this reason, a domain-specific tool should also offer the possibility to provide graphical DSLs.
In order to make modeling with graphical languages as intuitive as possible, a variety of operations must be provided that offer language-specific support in addition to pure editing.
In contrast to textual languages, however, the handling of graphical models requires the additional use of layout mechanisms to support the user in modeling.

\subsubsection*{Direct and Unrestricted Usability}

Before a tool can be used, a local installation is usually necessary.
Depending on how extensive the functionality of the tool is, it may be necessary to install additional runtime environments, SDKs or libraries.
In addition, there may be certain system requirements that restrict usability.
Thus, the necessary upfront work when using locally installed tools makes direct access difficult for a user.
This is in contrast to web-based tools, which can be accessed directly via the browser.
They do not require local installation, since all the necessary SDKs, libraries and runtime environments are prepared on a central server.
Thus, there are no special requirements on the resources of the system from the user, which allows unlimited usability on any devices with a browser.

\subsubsection*{Simultaneous Collaboration}

Collaboration between different users within a team is an essential component for an effective and productive working environment.
For this reason, collaboration should already be supported natively in a domain-specific tool.
The majority of domain-specific tools enable collaboration via version-control-systems (VCS) \cite{spinellis2005version}, which originate from computer science.
A VCS centrally manages the entire history and variants of the work state and allows users to import new versions as long as they do not conflict with the current state.
In this case, the user must manually resolve the conflict in order to propagate their changes.
However, this mechanism does not enable simultaneous collaboration in the form of concurrent collaboration in which the speed of change propagation plays the central role.
Similar to the established online office applications, especially non-technical users nowadays expect the possibility to work on a model or document simultaneously and in parallel instead of distributing it afterwards.

\subsubsection*{Immediate Executability}

In order to generate a decisive added value compared to a pure drawing and enable XMDD, it is necessary to bring the models to life as an instance of a DSL.
For this purpose, the associated tool must offer the possibility to execute a model.
The executability of models can be realized by three different approaches:
\begin{itemize}
\item Code generators are used to automatically translate a model into another mostly textual language (model-to-text) \cite{steffen2006model}. 
The generated code can be brought in a following step with the help of an external runtime environment to the execution.
\item Transformers differ from generators in that they do not generate code but another model (model-to-model) \cite{lybecait2018design}.
The produced model can be used for example as input of a further transformation or code generators.
\item Interpreters \cite{romer1996structure} allow to execute a model directly without conversion.
Here an execution environment already existing in the tool can be used, which traverses the elements of the model at the same time and executes a previously specified associated operation.
\end{itemize}
With regard to the required simplified usability for non-technical users, the execution of a model should be possible directly in the tool without additional installation and control of execution environments.
Immediate executability can accordingly be achieved with any of the aforementioned approaches, as long as any intermediate steps remain hidden from the user.

\subsubsection*{Specialized User Interface Design}

Besides the domain-specific language, the user interface (UI) of the enclosing tool is also a crucial factor for user acceptance \cite{10.5555/1197972}.
The UI of any tool or application is determined by its views, widgets and components.
A view describes an entire page or view on which different widgets are arranged.
A widget divides the view into different areas that serve a particular purpose.
Within a widget, different components are embedded, which represent control or display elements.

It is necessary that the UI is tailored to the respective target group by using common views, widgets and components.
In this way, the user is given intuitive access to the new tool, since he is already accustomed to the layout.
However, the familiar layouts differ, depending on the particular user group.
For example, programmers are used to tools that include a variety of widgets with components for version control, file explorer, outline, validation, TODOs, and so on.
This layout in turn appears cluttered and confusing to non-technical users, making intuitive access to the tool difficult.
For the target group of non-technical users, it must be possible to implement specific layouts, for example, to emulate views from LucidChart \cite{wwwlucidchart} and Miro \cite{wwwmiro}.

\subsubsection*{Management}

The established tools for local development such as Eclipse usually focus on supporting the individual user.
When working in a team, however, it is necessary to enable overarching management with regard to the role and authorization of the individual members.
Thus, it is common that depending on the respective expertise, certain groups and responsibilities arise.
In the area of domain-specific languages, this can mean that only certain users have the ability to use certain languages or parts of it.
These authorization mechanisms are usually not supported by the tool itself and can be mapped within limits using the VCS.
With regard to the One-Thing-Approach (OTA) \cite{ota}, the requirement arises to anchor this management within the tool.
Thus, the permissions and allocation of users to teams within an organization can be directly determined and enforced.

\section{Development challenges}
\label{sec:development_challenges}

In order to fulfill the described requirements efficiently, the use of language workbenches is indispensable \cite{fowler2005language}.
They provide support for recurring tasks in the development of domain-specific languages and the respective tools.
For the realization of a DSL, the definition of a metamodel is usually necessary, divided into the abstract and concrete syntax of the language.
Based on the metamodel, already rudimentary validation and support mechanisms can be generated automatically.
In addition, generators, transformers and interpreters can be implemented and integrated to make the DSL executable.
The requirements of the target group of non-technical users result in special challenges that have to be overcome in the development of DSL tools.

\subsubsection*{Metamodel and Instance Separation}

Since the users of a DSL should be able to work as easily and conveniently as possible with the respective tool, the instance and metamodel must be clearly separated from each other.
Even if tools such as MetaEdit+ \cite{tolvanen2003metaedit+} and WebGME \cite{maroti2014next} exist that allow the user of a DSL to modify it independently, it still contradicts the principle of simplicity by Steffen and Margaria \cite{margaria2010simplicity} and separation of concern (SoC) \cite{tarr1999n}, since additional and possibly unnecessary functionality is inserted.
In particular, non-technical users could be confused by the existing meta level, which in turn reduces acceptance.
Additionally, only from the clear separation arises the possibility to develop customized tools, instead of a single reflexive one, for the user.

\subsubsection*{Full Tool Generation}

In contrast to general-purpose languages, a DSL as described in the LDE approach \cite{steffen2019language} is tailored as precisely as possible to a specific target group and purpose.
This leads to the conclusion that a separate DSL must be provided for each target group and purpose.
In order to be able to do this within a reasonable scope, the creation of the DSL and the associated tool must be kept as simple as possible.
This applies in particular to the recurring work such as the declaration of the metamodel, validation and execution.
For this purpose, the concept of Full Tool Generation proposed by Naujokat et al.\cite{naujokat2018cinco} can be used to generate a domain-specific tool on the basis of a specification rather than developing it manually.

\subsubsection*{User Experience Specification}

In order to be able to create a tool that is as specific and customized as possible, the language workbench must also offer possibilities to influence the user interaction.
For this purpose, extension points should be available in the form of an event system, to react to the user's action.
Likewise a user must be able to be supported regarding the validity of a model and context actions.
Additionally introduced validators enable checks that do not directly emerge from the metamodel, such as model checking, data flow analysis and type checking, so that the user is guided to a correct model.

\section{DSL-Tool Development}
\label{sec:dsl_tool_development}

\begin{figure}
	\centering
	\includegraphics[width=0.45\textwidth]{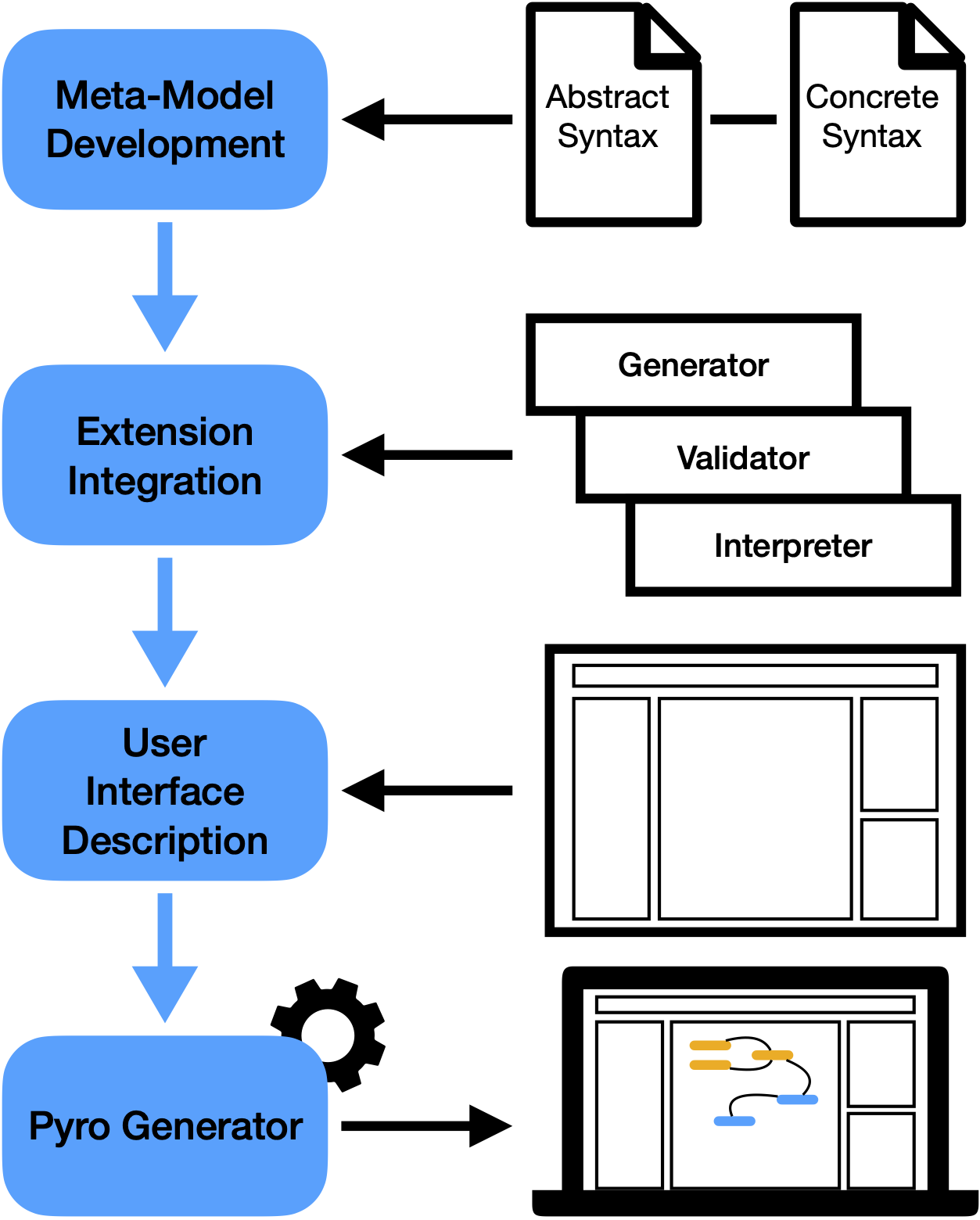}
	\caption{DSL development process using Pyro.}
	\label{fig:dev}
\end{figure}

The development of a new domain-specific tool poses many challenges (see section \ref{sec:development_challenges}), some of which can already be overcome at the meta-level:
\begin{itemize}
\item Definition of the graphical domain-specific language
\item Definition of the tailored user interface of the tool
\item Realization of the executability in form of a model interpreter
\end{itemize}
By using the generator integrated in Pyro, the complete domain-specific tool can be generated automatically based on these definitions. 
In this chapter, the individual steps of the high-level development process (see Fig. \ref{fig:dev}), which are necessary to create a DSL with Pyro, are explained.
The initial task of a tool developer is to define the domain-specific language by declaring the abstract and concrete syntax.
Pyro limits itself to graph-based graphical languages, so the underlying meta-metamodel of a DSL is already specialized and simplifies development.
Thus, according to Pyro's approach, each language consists of extensions of the meta-types: Edge, Node, and Container.
The resulting structure is defined in terms of connection and embedding constraints, which already allow the definition of static semantics.
In addition to the structure of the language, the representation plays a crucial role in providing users with intuitive access to the DSL.
The individual shapes and colors must be aligned with the established notation of the domain.

To enable users of the DSL to obtain something functional additional components must be created.
These components include translators in the form of generators, transformers and interpreters as well as validators to realize a real tool instead of a "smart paper".

Another decisive factor for user acceptance is the design of the tool interface.
Depending on the intended target group, familiar layouts and components should be used to enable intuitive and easy access.
This user interface design is already supported by Pyro on the meta-level.

Subsequently, the Pyro generator takes over, which translates the previously created metamodels within a generation step into an executable, web-based modeling environment.
This translation from the meta-level to the instance level is performed by various generators within Pyro, which generate a holistic distributed application based on the previously determined abstract and concrete syntax.
From this point on, users are able to use the newly created DSL embedded in a fully specialized tool from anywhere and without prior installation.

\subsection{Language Metamodel}

A metamodel combines the concepts and structures present in a concrete model in the form of a formalized description. 
In relation to the metamodel of languages, formal structures and composition possibilities are defined, which describe the abstract and concrete syntax.
In textual programming languages, for example, the definition of a metamodel can be done with a Bakus-Naur form (BNF) \cite{mccracken2003backus}.
The BNF describes a metamodel in terms of terminal and non-terminal production rules.

With respect to graphical languages, however, more intuitive definition languages are used.
Many different description forms from the Unified Modeling Language (UML) family or the Eclipse Modeling Foundation (EMF) ecosystem exist to define a metamodel for graphical languages.
However, these meta-modeling languages are not specialized for the definition of graph-based graphical languages, which unnecessarily complicates the description.

For this reason, Pyro is based on a specialized meta-metamodel that already includes the concepts of graph-based graphical languages. 
Thus, creating a new domain-specific language with Pyro is done by instantiating the meta-metamodel to define the abstract and concrete syntax.
To simplify the instantiation of the meta-metamodel, language workbenches such as CINCO or EcoreTools can be used.
CINCO is based on the same meta-metamodel as Pyro and has already been successfully used for the creation of various graphical DSLs \cite{10.1007/978-3-319-47169-3_60,10.1007/978-3-319-47169-3_58,gossen2018dsls}.
The advantage of CINCO lies in the existing own declarative DSLs for the specification of the metamodel of a language.
In the following sections, the meta-metamodel for the abstract and concrete syntax of Pyro is described.

\subsubsection{Abstract Syntax}

\begin{figure}
	\centering
	\includegraphics[width=0.45\textwidth]{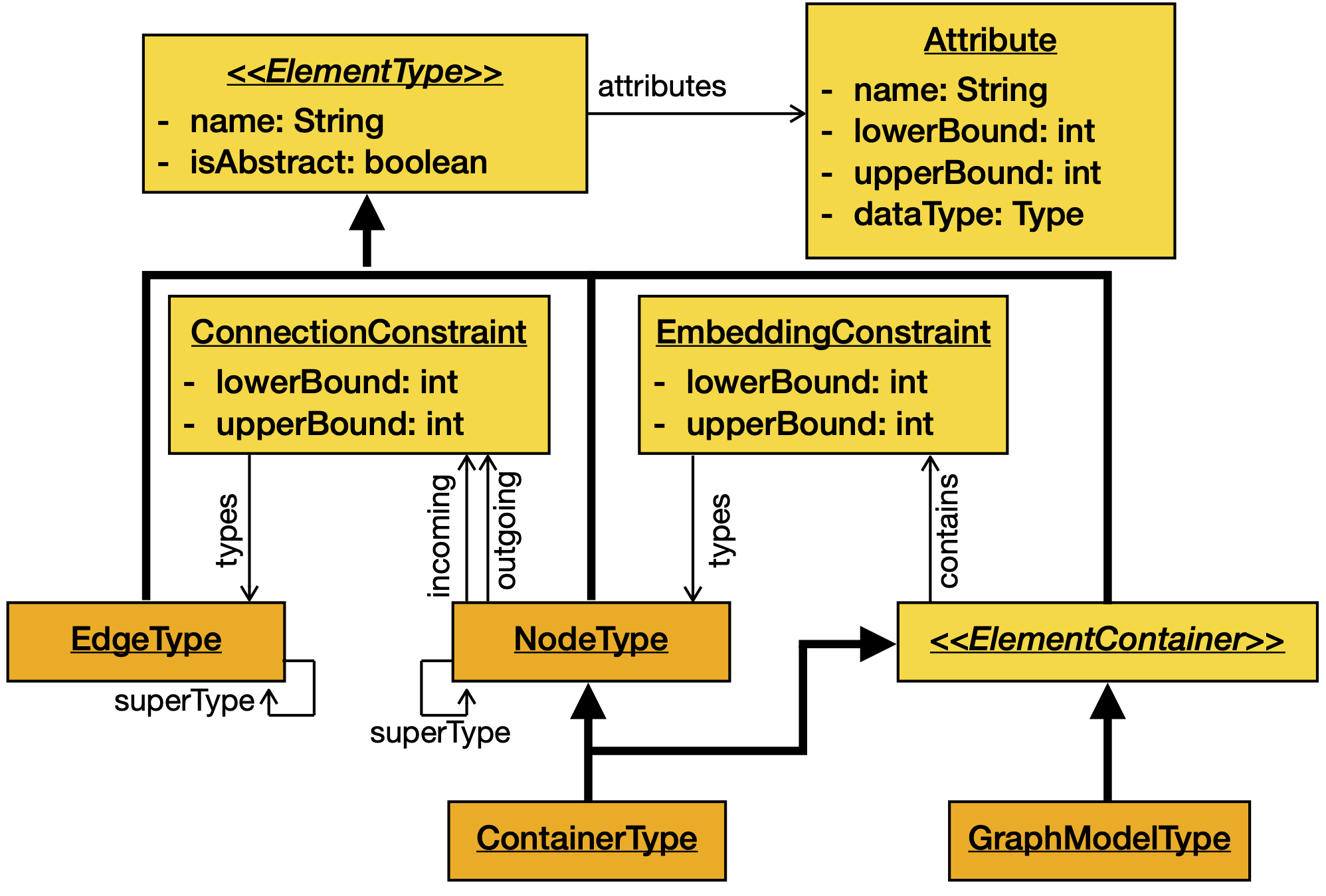}
	\caption{Meta-metamodel for the abstract syntax definition.}
	\label{fig:mgl}
\end{figure}

In the field of graph-based graphical DSLs, the abstract syntax represents the set of all element types available in the language.
However, by specializing in graph-based graphical languages, this set is reduced to refinements of the meta-types: Node, Edge, and Container, which must be defined in the context of a concrete DSL development.
Figure \ref{fig:mgl} shows the meta-metamodel of Pyro's abstract syntax in the form of a structure diagram.
A distinction is made between associations (thin edges) and inheritance (thick edge with filled tip).
The meta-metamodel contains different types which can be instantiated in the context of a DSL development.
The GraphModelType is the root of the definition and represents a concrete DSL.
As can be seen, the GraphModelType inherits from ElementContainer, which in turn has the possibility to reference different node types via the EmbeddingConstraint.
This way it is possible to specify which and how many nodes of a certain type can be included in the graphmodel.

The NodeType is used to specify an available node type in the graphmodel.
By the associated ConnectionConstraint type, it can be defined to a NodeType, which edge types can incoming and outgoing including the cardinalities.

The ContainerType inherits from NodeType and ElementContainer, combining the properties of both.
By inheriting from ElementContainer, it is possible to define which and how many nodes can be contained in the specific ContainerType.

As seen in Fig. \ref{fig:mgl}, the meta-metamodel allows the definition of polymorphism through the superType association on the NodeType and EdgeType.
In this way, an inheritance hierarchy of node or edge types can be defined to structure the elements of the language.
Similarly, there are possibilities for grouping types, which simplifies the definition of embedding and connection constraints.

The instance of an EdgeType describes an edge available in the language, where only the ConnectionConstraints of a NodeType can define where it can be drawn.

All base types present in the meta-metamodel extend the ElementType, which represents the commonality of all nodes, edges, containers, and the graphmodel type.
Thus, the ElementType defines that all instances are named and can be declared as abstract.
As known from the object orientation, an abstract type cannot be created in the described language and is only used for grouping.
In addition, any number of attributes can be defined for each ElementType, which allow the user of the language to set the respective values at modeling time.
Each attribute is also named, typed and given cardinalities.

Based on the meta-metamodel, an instance can be created that represents the graph-based language.
This instance can then be used as input for the Pyro generator to completely generate the described language including the complete tool.

\subsubsection{Concrete Syntax}

In addition to the declaration of which nodes, edges and containers will be present in a DSL, it must also be defined how the respective elements will be visualized for the user.
Thus, a corresponding style must be defined for each non-abstract ElementType.

The definition of visualization can take place in various ways and differs in each case strongly depending on the final execution environment.
In view of the intended simplicity of Pyro, the visual representation of an element is defined in terms of hierarchically arranged shapes, following the established hierarchical structure of SVGs \cite{eisenberg2014svg}. 
\begin{figure}
	\centering
	\includegraphics[width=0.45\textwidth]{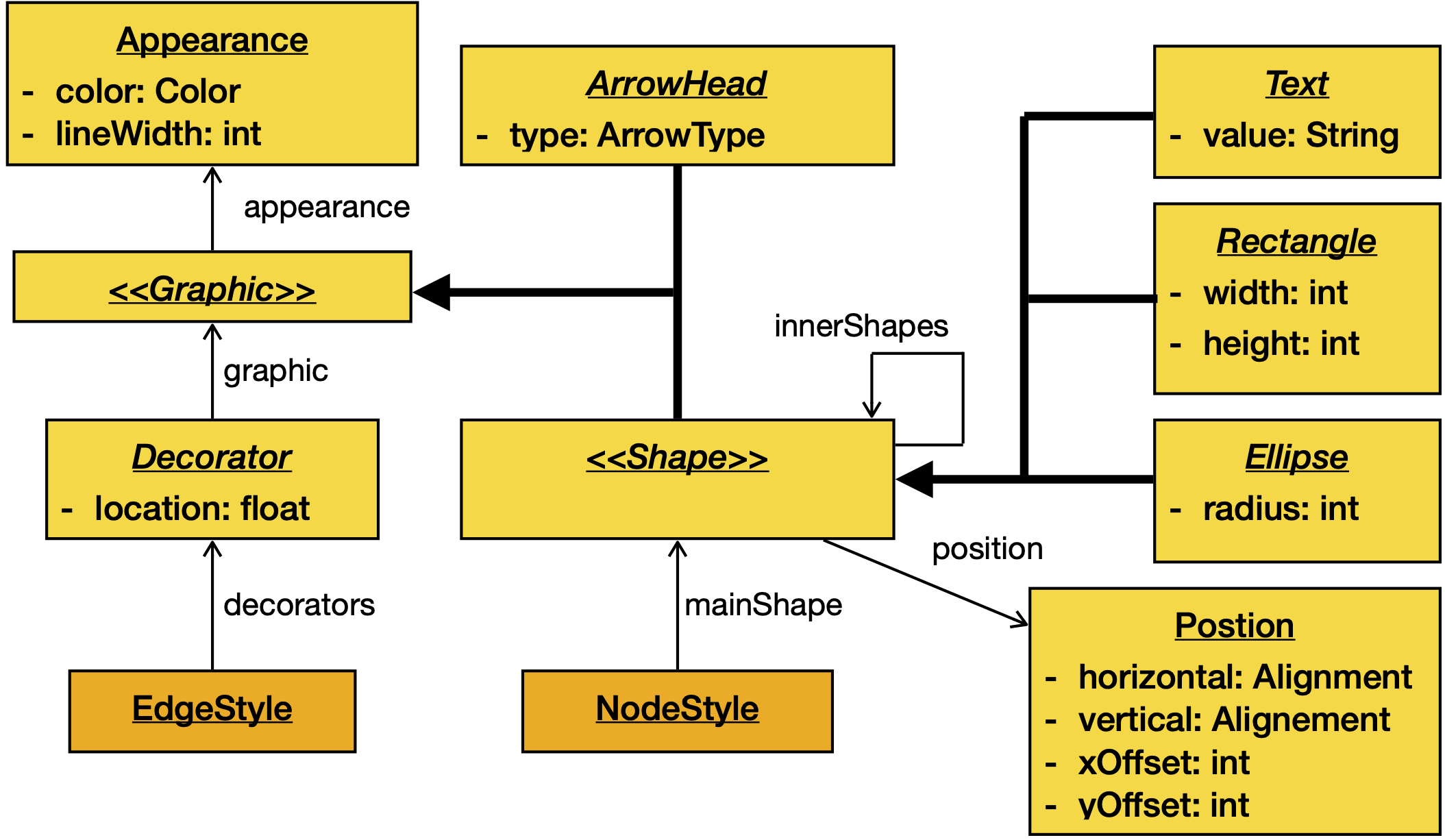}
	\caption{Meta-metamodel for the concrete syntax definition.}
	\label{fig:msl}
\end{figure}
For this purpose, Pyro provides another meta-metamodel (see Fig. \ref{fig:msl}) that can be instantiated to define a concrete syntax.
First, a distinction is made between the style definition of an edge and that of a node by instantiating either the NodeStyle type or the EdgeStyle type.
As before, the meta-metamodel includes thin edges for visualizing associations and thick edges with a filled tip for inheritance relationships.

The instance of a NodeStyle type describes the concrete visualization of an associated node.
The definition of a NodeStyle begins with the outermost Shape, which is represented by the mainShape association.
The associated Shape type is extended by different concrete shapes like Rectangle and Ellipse.
Each Shape type has special attributes to define properties related to e.g. width and height of a Rectangle.

To create complex styles, the shapes can be arranged hierarchically on top of each other.
The innerShapes association allows for example to place a Text shape inside a Rectangle.
In Fig. \ref{fig:msl} only a part of the Shape types available in the meta-metamodel is shown to illustrate the concept.

The positioning of the shapes among each other is done relative to the parent by first determining the horizontal and vertical anchor point and then specifying an offset.
The corresponding values can be stored via the attributes of the associated Position type.

In contrast to the NodeStyle, the EdgeStyle is not based on a single shape.
The visualization of an edge is determined by the definition of multiple, relatively positioned decorators.
For this purpose, the meta-metamodel provides the Decorator type, which can be positioned relative to the edge length via the location attribute.
A Decorator can be either a Shape or an ArrowHead, which is made clear by the extensions to the associated Graphic type.

In addition to specifying shapes, it is necessary to define colors and other subtleties.
For this reason, the meta-metamodel for the concrete syntax contains the Appearance type which is associated by Graphic and thus can be instantiated for any Shape or ArrowHead.
The Appearance type contains a variety of different attributes for defining colors, line thicknesses, fonts, and so on, which is only partially shown in Fig. \ref{fig:msl}.

Basically, Pyro pursues the goal of separation of concerns with the division into two meta-metamodels, which is why the visualization is defined separately from the structure.
This approach makes it possible to declare complex DSLs with a meaningful appearance in a simple and intuitive way.

\subsection{Executability}

Domain-specific languages offer users the possibility of independently mastering certain problems within their domain.
In the context of model-driven software development, a central goal is to generate a complete software system or parts of a software system from the models created with a DSL \cite{stahl2006model}.
In contrast to pure representation and design DSLs, an executable DSL offers the greatest added value for non-technical users to realize a program without programming knowledge and bring it to execution.
This approach of XMDD has been successfully used in many tools such as Bio-jETI \cite{lamprecht2009bio}, DIME \cite{10.1007/978-3-319-47169-3_60}, and jABC \cite{steffen2006model}.
The tailored abstract and concrete syntax of the DSL thus provides an intuitive approach to programming that is abstracted from the concrete technical realization.

The actual execution of a program described with the DSL can take place either by translation to source code of the target platform or by interpretation with the help of an execution model.
The translation of a model to a software system is done with the help of code generators, which traverse the individual elements within the model and successively generate associated source code \cite{czarnecki2000generative}.
For example, code generators are used to generate classes in an object-oriented target language (like Java or C++) from UML models \cite{parada2011generating}.
Subsequent to the generation, the code can be extended if necessary in a programming environment and finally executed.
However, the subsequent adaptation of the code results in round-trip engineering \cite{hettel2008model}, in that the source model diverges from the generated code. 
\begin{figure}
	\centering
	\includegraphics[width=0.45\textwidth]{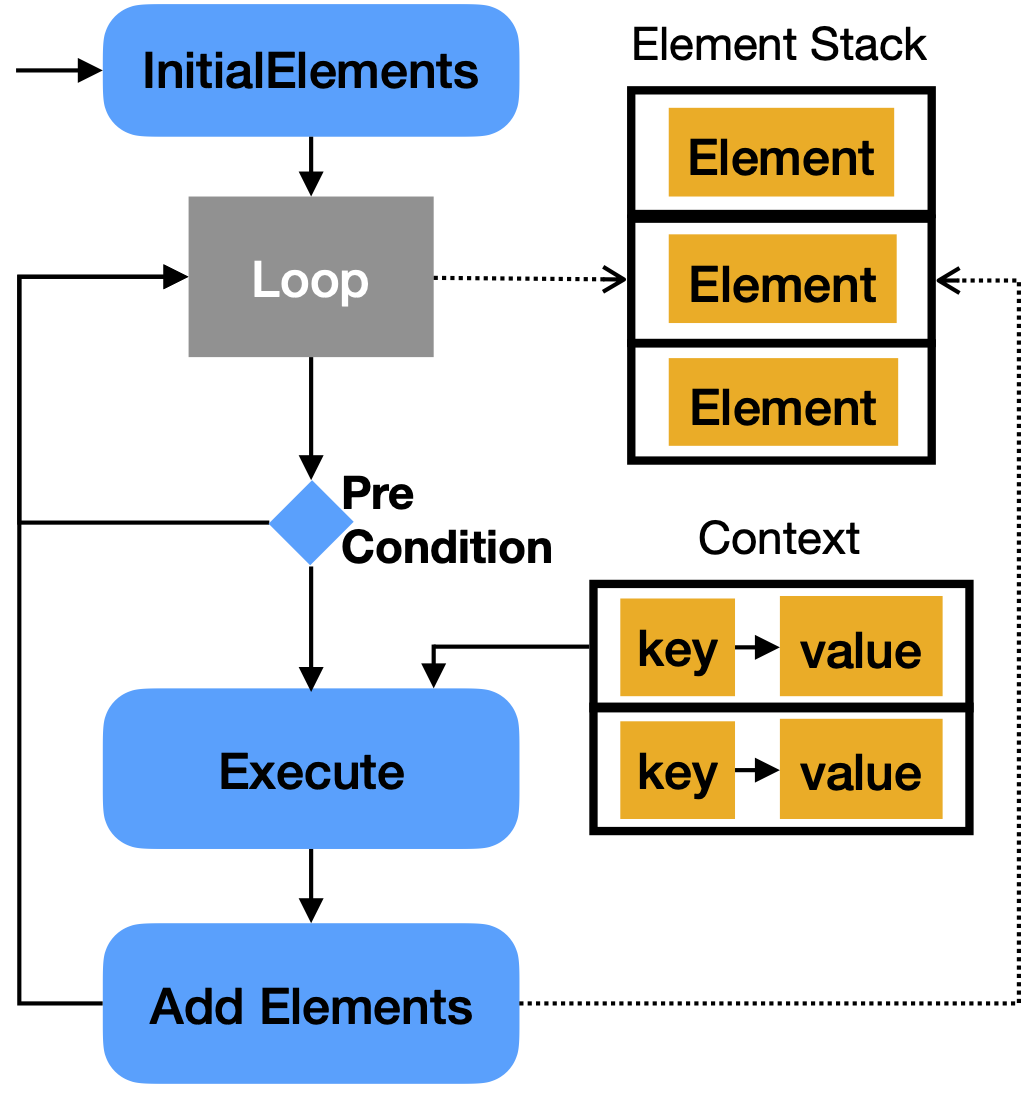}
	\caption{Interpreter framework execution model.}
	\label{fig:interpreter}
\end{figure}
As opposed to generating code, interpreters can be used to execute code or models seemingly directly.
The interpreter reads for this step by step the individual instructions in form of model elements, interprets it on the semantics and executes it in a background execution environment.
Compared with the execution of code generated before, interpreters are however 5 to 20 times slower \cite{waite2012compiler} in their execution speed, which is owed to the circumstance that the generated source code can be optimized before the execution by a compiler in contrast to the interpreter.
The advantage of interpreters lies however in the independence of the existing computer architecture, since the code which can be interpreted can be executed on each system, on which the interpreter is installed.

In terms of realizing executability as easily as possible, Pyro already provides a Xtend \cite{wwwxtend} framework for creating sequential interpreters for a graphical DSL defined via the meta-metamodel in the Java runtime environment.
The interpreter framework (see Fig. \ref{fig:interpreter}) is based on the approach of Voelter \cite{voelter2013dsl} by assigning an instruction to each node, edge or container type and executing it step by step.
The individual instructions must be programmed in advance within the interpreter creation in Java, whereby the respective typed method stubs are already generated by Pyro based on the DSL metamodel.

A dispatcher executed before each statement decides which statement to execute based on the type of the element read.
In addition, the framework allows the creation of preconditions, which can be used to decide at runtime whether a statement is to be executed or not.
Within the execution of an instruction the possibility exists to store data in a global context and to read afterwards.
Thus a data flow can be established between the different instructions.
In order to be able to read data afterwards, they are stored under previously determined identifiers.
These identifiers can be textual as well as an element of the model itself.

To determine the execution order, the Pyro interpreter framework maintains a global stack that lists all elements that are still to be executed.
When creating the interpreter, a function must be programmed that determines and returns the initial model elements.
From this, execution is started and the instruction for the first element in the stack is triggered.
In addition to the initial order, the framework offers the possibility to add further elements to the stack after each instruction.
In this way, for example, conditional branching can be implemented by placing the next model element determined by the evaluation on the stack.

The interpreters developed with Pyro's framework can then be directly integrated and used in the tool.
The framework simplifies the step from a purely representational DSL to an executable one, even if the support is limited to simple sequential interpreters in which there is a direct relationship between model elements and instructions.

\subsection{User Interface Design}
\label{sec:user_interface_design}

The user interface (UI) enables a human to interact with a machine or system \cite{stone2005user}.
With regard to the use of software, graphical user interfaces (GUI) are usually used instead of physical buttons that are displayed on a screen.
The purpose of a graphical user interface is to provide the user with the best possible access to a software.
The user interface design plays a crucial role in this process, as it should offer the user the functions he needs to realize his intention without overwhelming or confusing him.
These challenges have to be mastered within the interaction design (IxD) \cite{rogers2011interaction} in order to achieve a user experience (UX) that is as intuitive as possible.
As described by Rogers et al., IxD requires a precise knowledge of the respective target group in order to create a successful GUI.

The same is especially true for the creation of domain-specific tools, which are also developed for a specific target group.
However, the established language workbenches do not provide separate support for the provision of specific GUIs.
Usually, a language workbench uses the same user interface for both the developers of a DSL and the users, although they may be fundamentally different groups.
\begin{figure}
	\centering
	\includegraphics[width=0.45\textwidth]{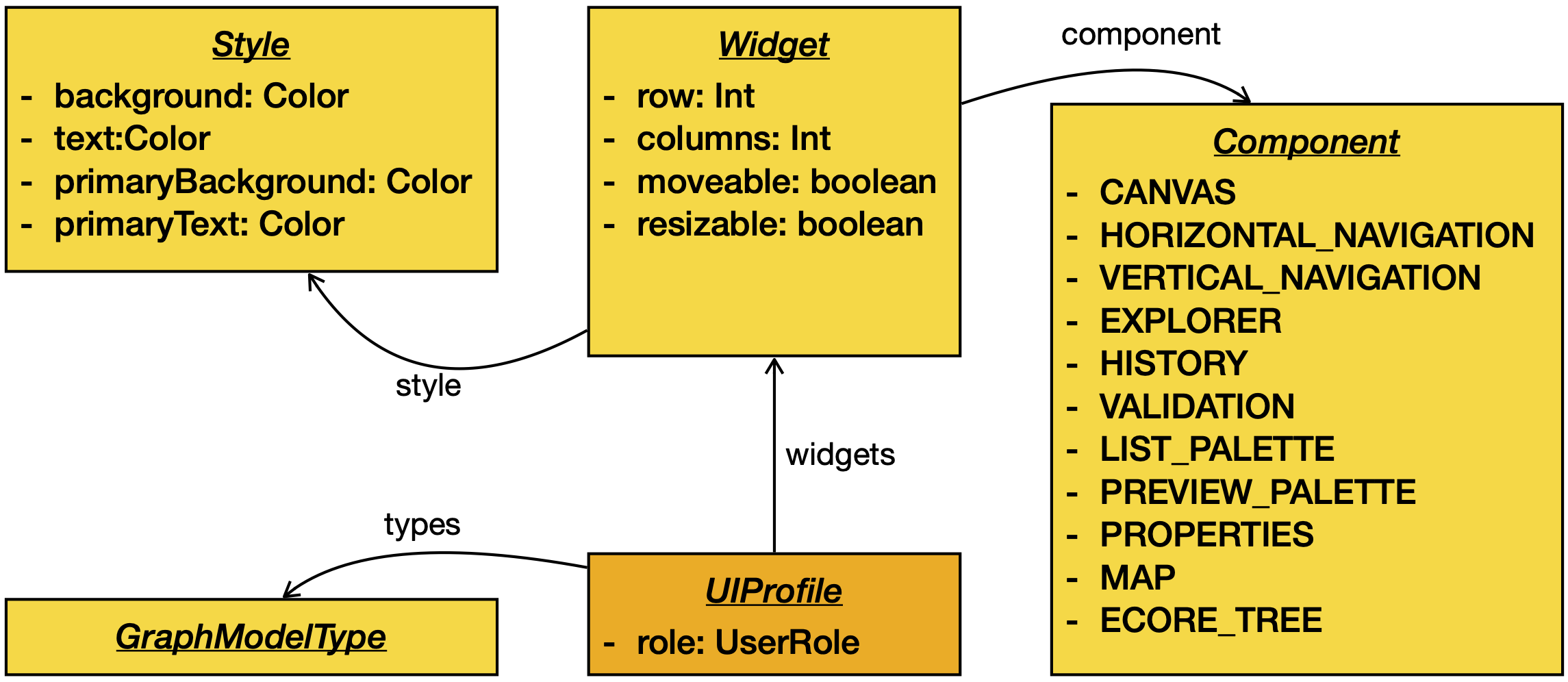}
	\caption{Metamodel for user interface specification.}
	\label{fig:ui}
\end{figure}
Pyro, on the other hand, tries to adapt the user interface of the domain-specific tool to the target group by providing separate support.
For this purpose, a separate metamodel (see Fig. \ref{fig:ui}) is provided, which can be used in parallel to the DSL development to describe user interfaces.
As before, associations within the metamodel are represented as thin edges.

The user-interface metamodel is based on the following approaches:
\begin{itemize}
\item A user interface can be associated to a specific set of user roles, so that different users can get different interfaces.
\item A user interface can be created for a specific set of the DSL within a tool.
\item The user interface combines a selection of specific visual components with an associated layout and styling.
\end{itemize}

As seen in the metamodel in Fig. \ref{fig:ui}, a particular user interface can be defined by an instance of the UIProfile type.
First, the UIProfile is assigned to a specific UserRole so that a specific interface can be created per user group.
A UIProfile consists of several associated widget types that can be arranged on the screen.

Pyro offers a grid-based layout, which is already used by established UI frameworks such as Bootstrap \cite{wwwbootstrap}.
The grid layout first divides a screen into an arbitrary number of rows.
Each widget within a row can specify a number of columns it should occupy.
The actual width of a widget is thus the total number of columns reserved in a row for all widgets.
For example, a widget with two columns in a row with a widget with six columns takes exactly one third of the available screen width.

In addition to the positioning determined by the rows and columns, the metamodel allows the specification of movement and magnification options of a widget.
In this way, a user could be given the ability to move widgets around the screen and resize them if desired.

The widget which will be displayed is specified by an associated literal of the Component enumeration.
Pyro provides a variety of components that can be selected to populate a widget.
Included are essential components like the CANVAS for modeling the graphical DSL and the VALIDATION for visualizing error messages but also special ones like the HISTORY component for chronological listing of modifications of a model.

In addition to the component, it is possible to define per widget how it should be colored.
For this purpose, the user interface metamodel provides the Style type, which has various color attributes to specify the appearance of a widget.

Instances of the UI metamodel can be used to tailor the user interface to different target groups.
It can be decided which components are required for a user group and how they are visualized and positioned to ensure intuitive access to the tool.

\section{System Architecture}
\label{sec:system_architecture}

The architecture decides which characteristics a software system fulfills and thus which possibilities a user has in handling the system.
From the requirements of the users formulated in section \ref{sec:user_requirements} certain characteristics have to be realized for an appropriate graphic modeling environment as software system:
\begin{itemize}
\item Due to the requirements of direct accessibility from everywhere, the necessity of a distributed system as a client-server application arises.
\item The multiple interaction possibilities in the graphical modeling can be achieved by using a rich client, which is realized as a single-page application \cite{gavrilua2019modern}.
\item Simultaneous collaboration can be achieved by using bidirectional communication channels that propagate the changes between server and client.
\end{itemize}
The elaborate realization of the targeted software system is usually the responsibility of the software developers and requires the use of diverse methods and frameworks.
By restricting to graph-based tools described via the meta-metamodels shown in section \ref{sec:dsl_tool_development}, the full generation approach of XMDD can be applied.
Pyro follows this approach by generating the entire distributed software system starting from the metamodel of the DSL and the UI.

The tool generated by Pyro consists of different components which are divided between the server and the rich-client which are implement by the Model-View-Controller (MVC) \cite{leff2001web} pattern.
The server manages the central instance of a model within a database.
To effectively store the instance of a DSL, Pyro generates a specific schema based on the metamodel.
A client accesses the model through the controller within the server.
The controller must, on the one hand, maintain the correctness of the model with respect to the static semantics of the metamodel and, on the other hand, resolve potential conflicts due to concurrency.

The client generated by Pyro includes rich capabilities to edit a model and send the corresponding modifications.
At the same time, the client must receive and display the simultaneous operations of other users.
In order to make the use of the client equal to that of a desktop application, the background synchronization must not interfere with the modeling.

The following sections describe which components are used within the generated system to meet the requirements.
In particular, it is explained how the individual components are generated depending on a concrete language.

\subsection{Model Layer}

\begin{figure}
	\centering
	\includegraphics[width=0.45\textwidth]{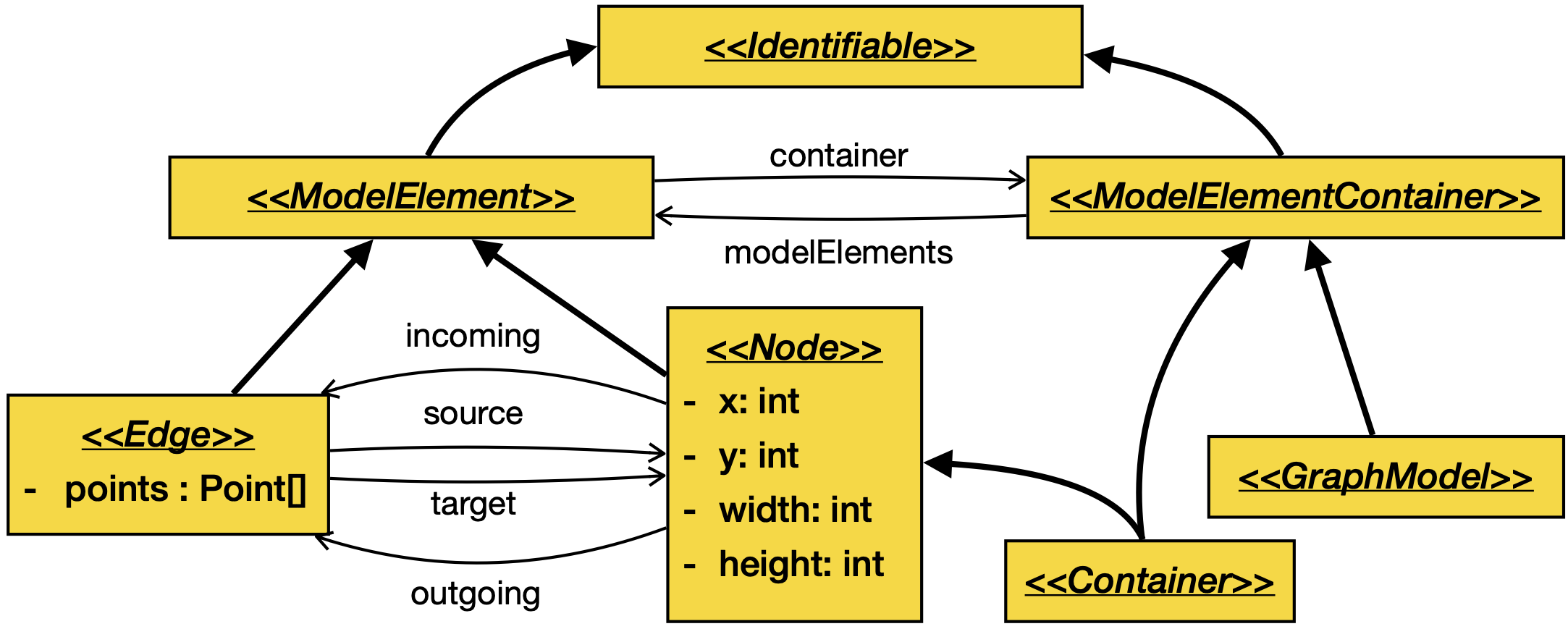}
	\caption{Common metamodel for a stored instance of a graphical DSL.}
	\label{fig:model}
\end{figure}

The central object to be stored within a graphical modeling environment is the model as an instance of a previously designed DSL.
Within the distributed system generated by Pyro, the model is persisted and managed in a central database on the server.
Since Pyro is limited to graph-based graphical languages only, each model is based on the metamodel of a graph.
Fig. \ref{fig:model} shows the common metamodel for persisting all graph-based languages, which is extended by additional nodes, edges and container types according to the DSL definition.
To store an instances of the shown metamodel in a database, a corresponding schema has to be designed.
The definition of a schema can be done by using entity relation (ER) models \cite{chen1976entity}.
ER models are used for the abstract specification of relational database schema that structures data in tables.
Additionally, rows within a table can be referenced by keys to implement relations.

The transformation of an ER model to a database schema can be realized by an Object-Relational-Mapper (ORM) \cite{o2008object}.
An ORM provides the bridge between a relational database and a metamodel in the object -oriented context by creating an abstraction layer between the object and the actual tabular persistence.
The ORM includes inherent rules on how the instance of a metamodel is stored in the database by translating the types, associations, inheritance relations, and attributes defined in the metamodel into a relational database structure.
Pyro generates annotated Java classes as input to the ORM.
There are several strategies how a metamodel can be mapped by the ORM into a relational database.
The Hibernate ORM Panache \cite{koleoso2020quarkus} used by Pyro is based on the following rules:
\begin{itemize}
\item A table is associated with each non-abstract type
\item For each attribute of a type a corresponding column is created in the table
\item Association is mapped via relations between the tables of the types. Depending on the cardinalities of the associations, additional relation tables are created.
\end{itemize}
As shown in the meta-metamodel shown in Section \ref{sec:dsl_tool_development}, an inheritance hierarchy can be declared between the individual nodes, edges and container types.
Inheritance between types is a particular challenge and can be realized by different strategies of the ORM, which differ in terms of performance and redundancy.
With respect to the metamodel of graphs in Pyro, these strategies were compared:
\begin{itemize}
\item \textbf{Single Table:} all classes that are in an inheritance hierarchy are persisted in a common table that contains all attributes of all classes.
In terms of the metamodel of graph-based languages (see Fig. \ref{fig:model}), all nodes, edges, and containers would thus reside in a single table, since they all inherit from the common super type: Identifiable.
This results in an inefficient data schema, in terms of a large number of unused cells and access times.
\item \textbf{Joined Table:} Each object is persisted split along the inheritance hierarchy by creating a table for each type in the hierarchy.
To read a complete object, a join must be calculated along the hierarchy between all tables involved.
Thus, efficiency decreases with increasing inheritance depth.
With regard to the DSLs focused by Pyro, this strategy is accordingly unreasonable.
\item \textbf{Table per Class:} Each object is persisted in its own table including all inherited attributes.
Thus, a complete object can be read directly from a table.
However, this strategy requires all polymorphic attributes to be rolled out along the possible sub types, which means an enormous development effort.
However, thanks to Pyro's full generation approach, this can be automatically generated directly based on the language's metamodel.
Thus, this strategy provides the most efficient persistence, as no unnecessary join operations are needed to read and no unnecessary columns are present per table.
\end{itemize}
\begin{figure*}
	\centering
	\includegraphics[width=0.9\textwidth]{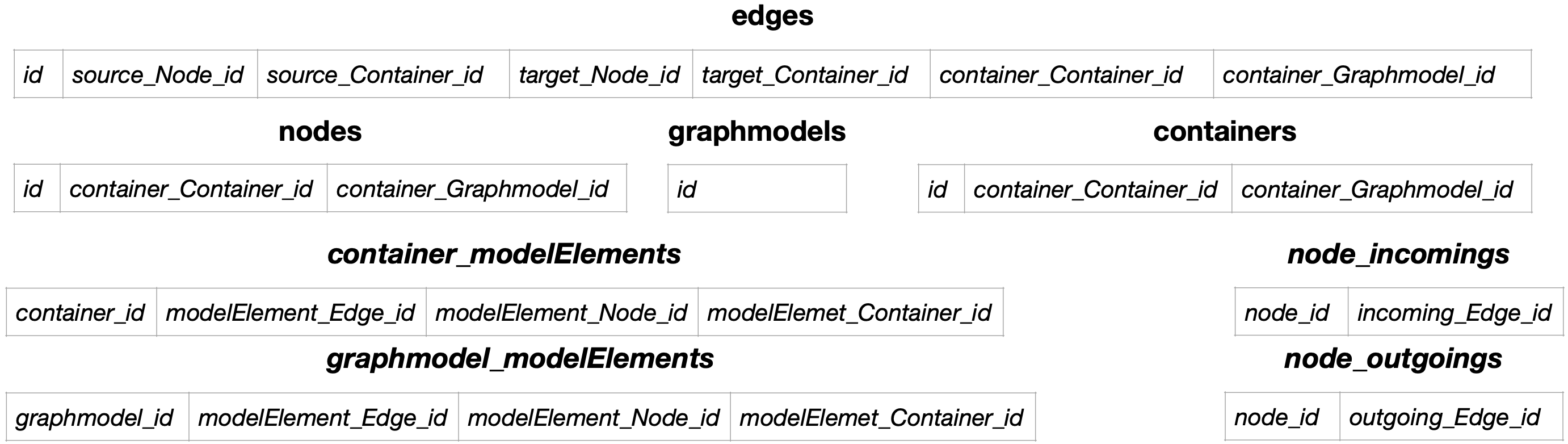}
	\caption{Generated database schema example for the metamodel by the ORM.}
	\label{fig:schema}
\end{figure*}
By using the join per class strategy, the ORM specifications can be generated by Pyro.
Fig. \ref{fig:schema} shows the database schema generated by Pyro and subsequently created by the ORM based on the metamodel.
As specified by the ORM strategy, each type within the metamodel is represented by a table.
Each attribute is represented as a column in whose cells the actual values can be stored.
Associations between the types are mapped by relations that reference the primary key of the respective row.

A special feature is the mapping of polymorphic associations like container between a ModelElement and the ModelElementContainer type.
Because of the inheritance hierarchy the target of the container association can be not only ModelElementContainer but all sub types like GraphModel and Container.
Since a relation within a database is always exactly one target table, the container association must be duplicated along the possible target types.
Therefore, as shown in Fig. \ref{fig:schema}, several relation columns are created for the container association - one for each possible target type.

By using the specialized ORM generated by Pyro for a language, an efficient model persistence can be realized in terms of query time and memory space, which is used by the controller.

\subsection{Controller Layer}

\begin{figure}
	\centering
	\includegraphics[width=0.45\textwidth]{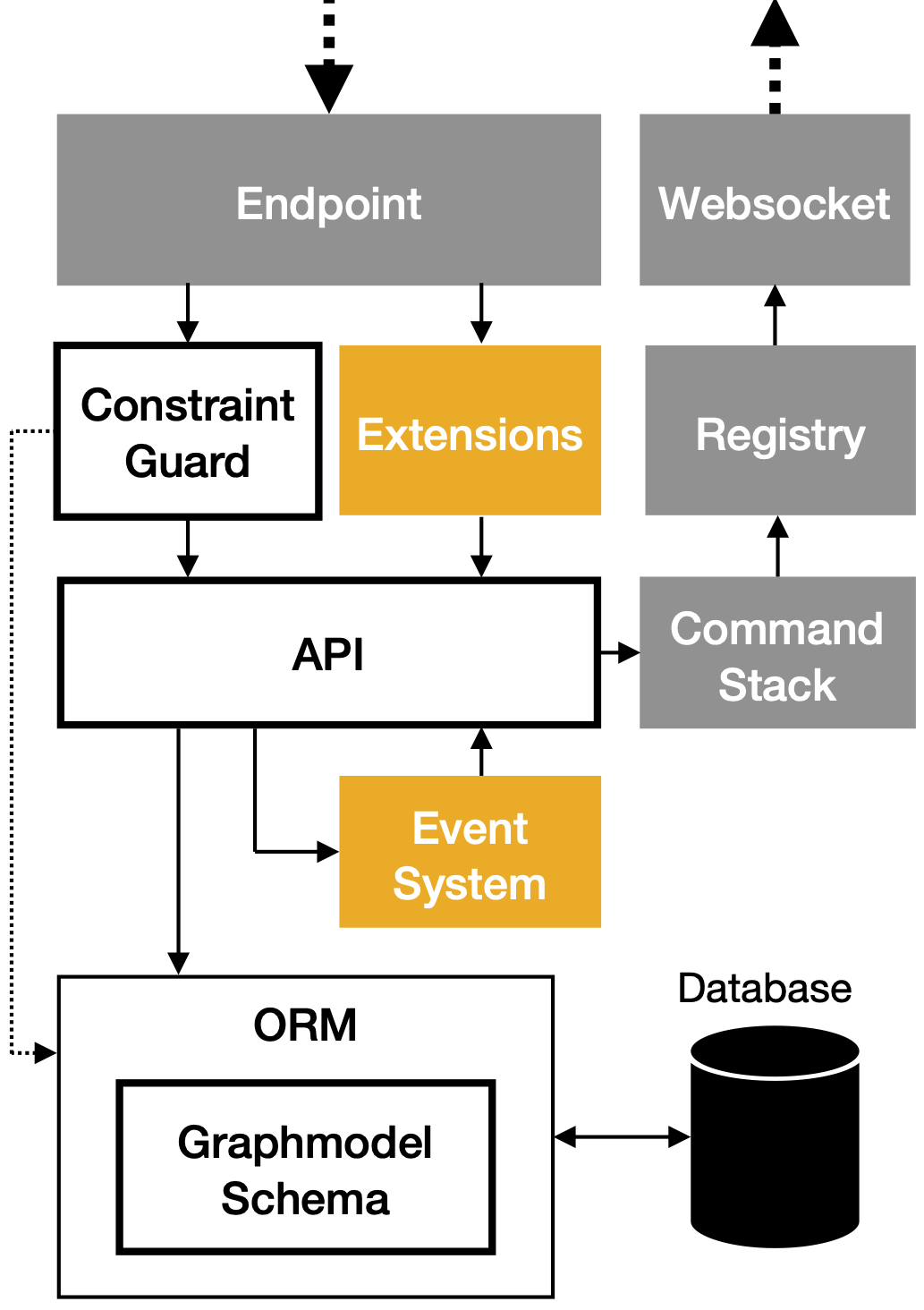}
	\caption{Sever application architecture showing generated (white), manually implemented (yellow) and static components (gray).}
	\label{fig:control}
\end{figure}

The controller layer of the system generated by Pyro serves as an access layer for the various clients to modify a centrally managed model.
The previously formulated requirements for the system result in certain properties that must be realized in the controller layer:
\begin{itemize}
\item The controller realizes a reactive application \cite{salvaneschi2013towards} by managing the bidirectional communication channels to the clients.
\item The syntactic correctness of a model must be controlled with respect to the static semantics defined in the metamodel.
\item The previously created extensions like interpreters and validators have to be integrated and controlled.
\item The conflicts that can result from the simulated modifications of the different clients have to be detected and managed.
\end{itemize}
As shown in Fig. \ref{fig:control}, there are several components within the control layer.
First, starting from the metamodel, Pyro generates a specific API layer that provides object-oriented access to the models stored in the database using the ORM.
The implementation of the API includes the Command Stack, which sequentially keeps track of all valid operations performed via the API.
The Constraint Guard is generated completely on the basis of the constraints defined in the metamodel of the language for embedding and connecting model elements.
It checks for each modification of the model initiated by the Endpoint whether the defined cardinalities are violated and thus prevents illegal operations.
If the operation was not rejected by the Constraint Guard, it is stored in serialized form in the Command Stack and then applied to the model.
The operations stored in the Command Stack are then distributed to all clients currently present in the Registry.

The Extensions possible in Pyro in the form of validators, interpreters, generators, context menu actions and the Event System all work on the generate specialized API.
In this way, it can be ensured that the model remains valid even for the execution of these integrated functionalities and that any changes to the model are recorded and subsequently distributed.
A special feature is the Event System, which offers possibilities to react to previously defined modifications of a model.
For this reason, the Event System itself is woven into the API and can nevertheless act exclusively on objects of it.

For communication between clients and the server, the controller layer provides a Websocket that enables bidirectional communication.
The Websocket manages all existing communication channels in a Registry for all clients working on the same model.
The valid operations of a client are propagated over the Websocket channels from the server to all other clients, thus achieving simultaneous collaboration between users.
The exact communication procedures, especially in case of conflict, are explained in Section \ref{sec:simultaneous_collaboration}.

\subsection{View Layer}
\label{sec:view_layer}

\begin{figure}
	\centering
	\includegraphics[width=0.45\textwidth]{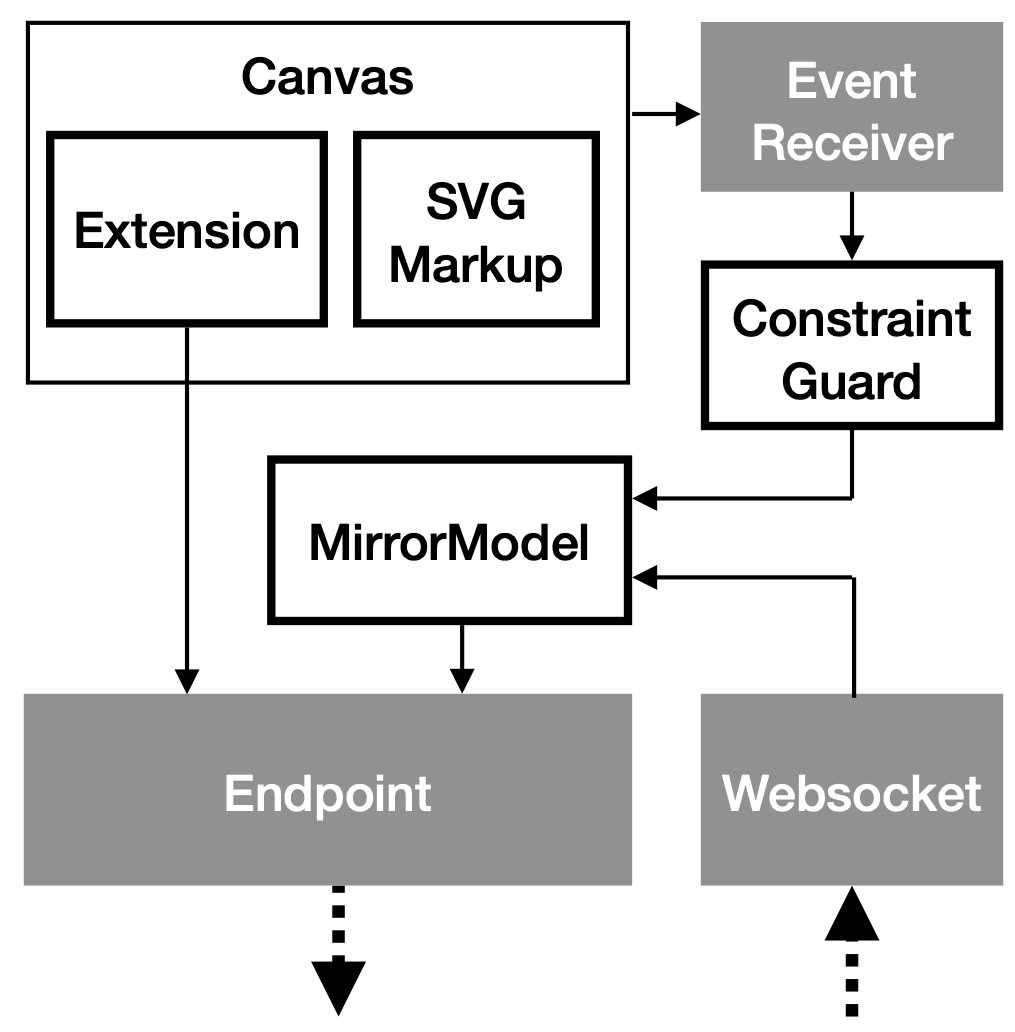}
	\caption{Rich-client architecture showing generated (white) and static components (gray).}
	\label{fig:view}
\end{figure}

The user acceptance of a domain-specific tool depends mainly on the user interface, the DSL and the interaction design realized in it.
In order to achieve the highest possible user acceptance, the client generated by Pyro must fulfill various requirements that are reflected in the system design:
\begin{itemize}
\item To ensure uninterrupted processing of a model, the client must be able to act as independently as possible.
\item The editing of a graphical model must be intuitive and direct via a canvas.
\item The client must reflect the previously defined user interface. 
\end{itemize}
As shown in Fig. \ref{fig:view}, the rich client generated by Pyro consists of different components to meet the formulated requirements.
Technically, the client is based on the Angular framework \cite{wwwangular} which, due to its single-page properties, offers a wide range of UX possibilities that are in no way inferior to those of a desktop application.
The independence of the client is achieved by the so-called MirrorModel, which is initially loaded from the server.
The MirrorModel is a complete copy of the centrally managed model and allows the client to make modifications independently.
The corresponding code for the management of the MirrorModel is generated by Pyro based on the metamodel of the language.
This way it is already possible on the client side to validate the constraints of the metamodel locally and to give direct feedback to the user without additional communication to the server.
Of course, all locally valid operations still have to be transferred to the server and validated against the central state to enable collaboration.

Starting from the loaded MirrorModel, the graphical representation is generated based on the concrete syntax.
For the visualization and editing of a model, Pyro provides a rich graphical modeling environment based on drag'n'drop as an interaction mechanism.
For this purpose the framework JointJS \cite{wwwjointjs} is integrated which allows the interactive manipulation of SVGs in the browser.
The SVG markups required by JointJS are generated based on the concrete syntax so that each element type receives a corresponding graphical representation.

User modifications to a model are detected by the associated Event Receiver, validated locally by the Constraint Guard, and applied to the MirrorModel.
To enable simultaneous collaboration, each operation performed by the user is serialized and transmitted to the server via the Endpoint.
Similarly, the Websocket is also used to receive other users' modifications and apply them locally.
The individual commutation steps within the collaboration will be explained in detail in the next section \ref{sec:simultaneous_collaboration}.

In addition to the Canvas, the client provides other components such as the palette, validation, or history, which are included and arranged depending on the user interface definition.
For this Pyro generates a special editor view for each defined UIProfile (see section \ref{sec:dsl_tool_development}) which embeds the selected components at the defined position.
Each component can access the MirrorModel to fulfill its respective purpose and, if necessary, fetch additional information from the server.

\section{Simultaneous Collaboration}
\label{sec:simultaneous_collaboration}

Collaboration is an essential requirement for a web-based domain-specific intuitive modeling environment as described in section \ref{sec:user_requirements}.
It allows different users to collaborate simultaneously on the same model.
Especially for Pyro's target group of non-technical users it is important that collaboration is possible without any manual synchronization of a classical version control system.
Even if this means that possibilities such as branching and a versions history are lost.

However, the realization of simultaneous collaboration poses several conceptual and technical challenges that need to be overcome:
\begin{itemize}
\item The system must be scalable with respect to a large number of users acting in parallel.
\item The modifications of a user have to be detected and propagated.
\item The propagated modifications have to be compared, validated and synchronized centrally.
\item The detected differences to the central model must be automatically distributed between the users to create an overall consistency.
\item Possible temporal and syntactic conflicts must be detected and resolved.
\end{itemize}

\subsection{Communication Model}
\begin{figure*}
	\centering
	\includegraphics[width=0.9\textwidth]{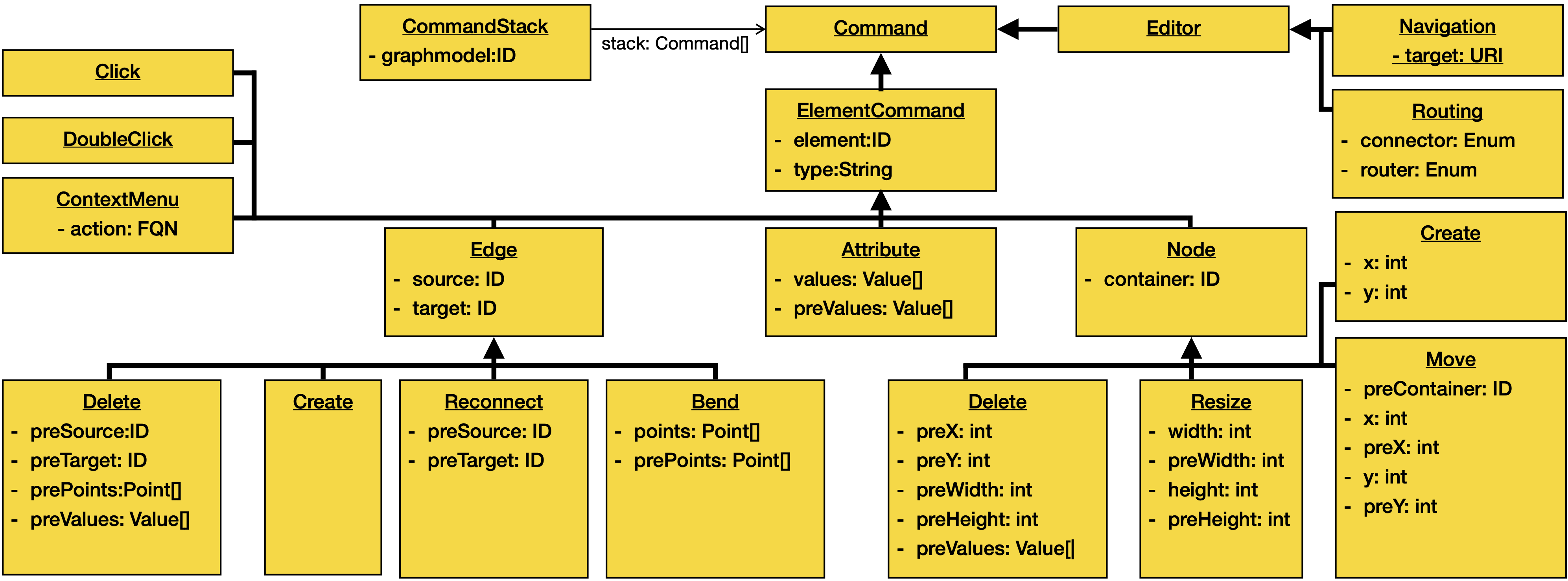}
	\caption{The message metamodel to realize the CRDTs by different command types.}
	\label{fig:messages}
\end{figure*}
To meet these requirements, the modeling environment generated by Pyro implements the optimistic replication strategy \cite{saito2005optimistic} by providing a duplicate of the model to each client.
The concept allows the respective duplicates to diverge for a short moment, which is achieved in Pyro's architecture by the MirrorModel on the clients (see section \ref{sec:view_layer}).
Thus, the consistency model implemented by Pyro belongs to the strong eventually consistent class \cite{vogels2009eventually}, which satisfies the properties basically available, soft state and eventually consistent (BASE) \cite{pritchett2008base}.
These properties allow for high availability by only requiring operations to be distributed on the elements of the MirrorModel.
With respect to Pyro's graph-based languages, these operations are limited to all possible modifications to nodes and edges.

To realize simultaneous collaboration and conflict resolution, a user's operations are serialized in the form of conflict-free replicated data types (CRDTs) \cite{shapiro2011conflict}.
CRDTs were originally designed for the synchronization of textual languages and simplified operational transformation \cite{sun1998operational}.
The synchronization of textual languages poses a particular challenge in determining the identities of individual text fragments.
The CRDTs extend the approach of operational transformation by creating an identifier over the responsible client , the changed value of a character and the position in the text.

In the area of textual languages, the Language Server Protocol (LSP) \cite{bunder2019decoupling} established itself on this basis for synchronization between client and server.
However, since textual languages differ fundamentally from graphical ones in their structure, visualization and way of editing, it is impractical to choose the LSP as a basis \cite{rodriguez2018towards}.
For this reason, the Eclipse Graphical Language Server Protocol \cite{wwwglsp} was developed, which is explicitly tailored to the graph-based structure and possible operations.
However, the GLSP is based on the concept of a thin client which cannot act independently in any way and thus offers only low availability.
Furthermore, GLSP does not support simultaneous collaboration, which is an essential user requirement.

Pyro complements the concept of GLSP in terms of collaboration by allowing CRDTs to be propagated through bidirectional communication channels.
Similar to GLSP, the CRDTs in Pyro will be realized by element-specific commandos that reflect the possible modifications.
Each command unifies the executed operation, the identifier of the element, and the previous state of the element to enable rollback, undo, and redo functionalities.
Thus, Pyro's CRDTs combine an operation-based modification distribution with a state-based addition for consistency checks and reversions.
Thanks to the use of CRDTs, conflicts between parallel modifications can be detected centrally.

\subsection{Command Types}
\label{sec:command_types}

The CRDTs used by Pyro describe the atomic modifications of a model by a user, which are serialized in the form of commandos and transmitted to the server as a message.
A message contains information about the sender and the respective modification, which is described by the message metamodel in Fig. \ref{fig:messages}.
The metamodel of a message shows the different types that are instantiated by Pyro to distribute the operations between client and server.
The messages are used within the bidirectional communication channels for both directions, so that both client and server provide the same interface.
Other communication channels exist to load additional information from the server such as validation results.
However, these communication types are not explained here because they have no influence on the collaboration between the users and cannot violate the consistency.

The root type Message contains the information which model was edited and which user performed the modification.
The respective element specific commands are serialized as an ordered list of Command type instances.
Depending on the type of operation, the corresponding Command type is instantiated and added to the list.
First, a distinction is made between Element and Editor specific commands.

Editor specific commands can for example mean a change of the used edge layout, which is represented in the form of a Routing command type.
Pyro supports different layouts which are composed of a specific connector and routing algorithm.
The connectors decide how the bending points of an edge are represented and routing algorithms decide how the edge is routed between nodes.

The element specific commands encode the modification on the canvas.
A distinction is made between nodes, edges, and attributes.
An edge can be created, deleted, reconnected, or bent.
Depending on the respective operation, the necessary information is serialized in the command.
For example, when an edge is created, the corresponding start and target nodes are specified by their IDs.
Additionally, information is stored that can be used to undo a command afterwards, such as the previous start and target nodes in a reconnection operation.

Nodes can also be created and deleted, but they are always placed inside a container or the graphmodel which is also stored in the command.
This information is also needed when a node is moved, because the container of the node might have changed by the movement.
For the change of the size of a node the Resize command is used which serializes accordingly the old and new size.

Besides the specific node and edge command types, the attributes of the elements within the client can also be changed.
For this purpose, the Attribute command type is used, which serializes the previous and new assignment of all attributes and thus transmits the modification to the server.

In view of the event system present in Pyro, it is also necessary to react to various user interactions that do not cause a direct modification of the model.
For this purpose, the command types Click, DoubleClick and ContextMenu are used, which can subsequently lead to a modification of the model on the server, if a corresponding extension has been integrated.

\subsection{Communication Process}
\label{sec:communication_process}

The communication process of Pyro describes how the individual messages of a client are received by the server, processed and then distributed.
First, a client loads the current state of the centrally stored model after it has been started and saves it as a local MirrorModel.
As soon as a modification of the model has taken place on the client, a corresponding message with the respective command is created and sent to the server.

\begin{figure}
	\centering
	\includegraphics[width=0.45\textwidth]{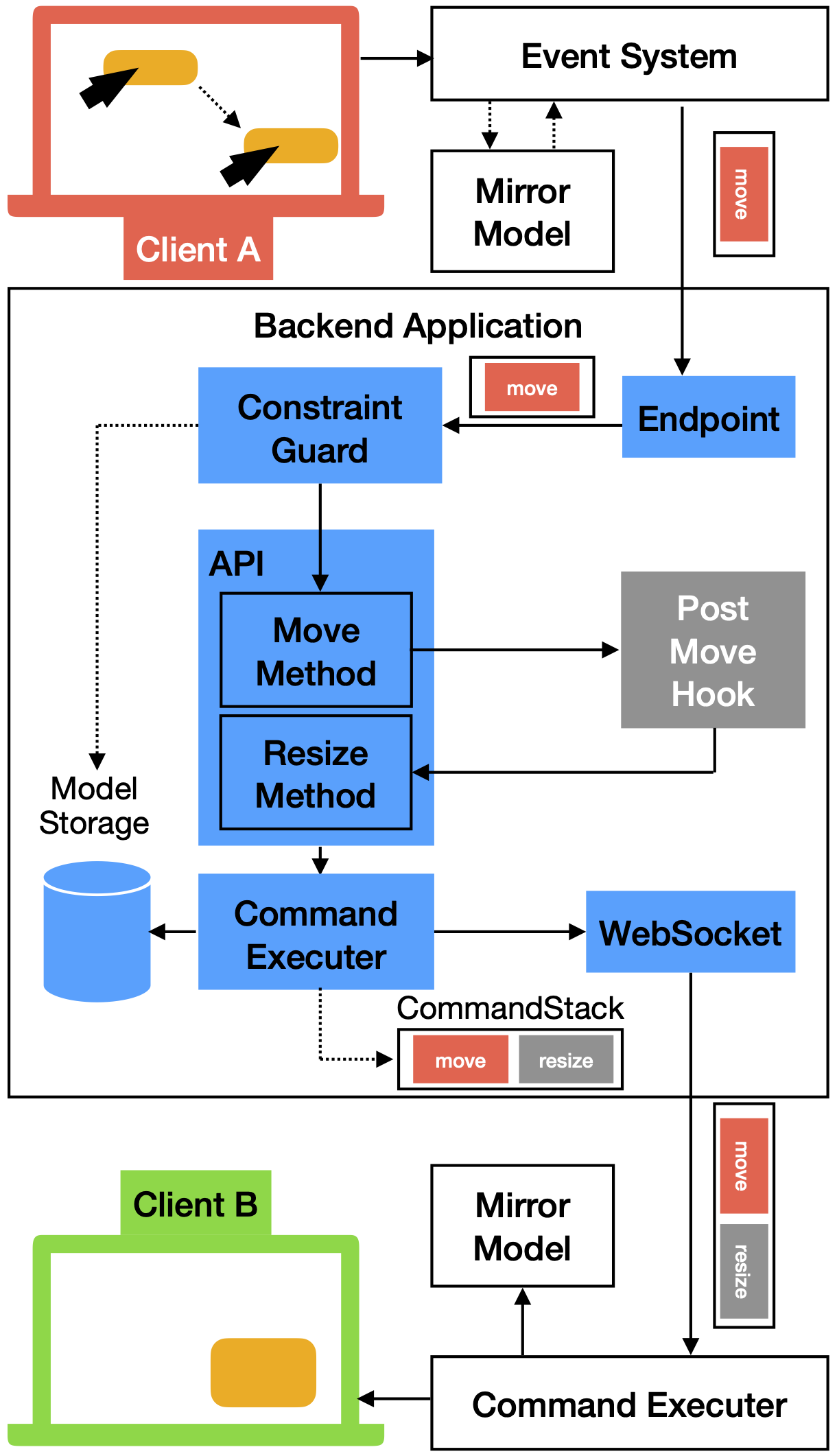}
	\caption{Exemplary communication process for the movement of a node with an server side event system extension.}
	\label{fig:comm_process}
\end{figure}

Fig. \ref{fig:comm_process} shows the process and the involved components of client and server when processing and distributing a received message using the example of a node movement with an event system extension.
In this example, the extension listens for the movement of a node and performs a subsequent resize operation of the node.

As soon as the user of client A moves a node by drag'n'drop on the canvas, the event system behind is notified.
The event system then uses the MirrorModel to check whether the constraints defined by the metamodel of the language are violated.
If the user's change is valid on the local state of the MirrorModel, the change is applied locally, serialized as a command, and finally sent as a message to the server.
The message contains information about the sender as well as the type of operation performed, as described in section \ref{sec:command_types}.
In this example, a Move command would be included in the message, encoding the node's ID, previous and new position.
The server receives the message as a request via the endpoint and first checks whether the message complies with the underlying protocol.
Then a transaction is started so that all subsequent changes to the database are atomic.
Based on the transmitted graphmodel ID, the model is loaded and the corresponding node is determined.
If both were found, the API for the node can be instantiated and the corresponding move method is called with the passed parameters.

Within the API, the ConstraintGuard first checks whether the constraints specified via the metamodel are violated with regard to the centrally stored model.
If the intended operation is allowed, the called move method is executed by calling the underlying Command Executer.
The Command Executer is the link to the ORM and applies the changes to the database.
In this case the x and y coordinates of the node are adjusted according to the passed parameters.
At the same time the Command Executer manages the CommandStack which keeps track of all valid modifications within the current transaction. 

After the move operation has been applied by the Command Executer and tracked in the CommandStack, Pyro's event system is triggered, which in this example invokes an extension of the move method.
The corresponding Post Move Hook is called and processed with the newly positioned node.
In this example, the Post Move Hook is used to subsequently resize the repositioned node.
The Post Move Hook uses the same API so that, as before, the operation is first checked, applied and tracked in the CommandStack.
Thus the CommandStack now contains both the move operation and the enlargement of the node.

Finally, the CommandStack is distributed to the initial client (not shown in Fig. \ref{fig:comm_process})  and all other listening clients via the WebSocket.
In this example, a client B is listening which receives the two commands in a message from the server.
The received commands are applied directly to the MirrorModel of the client.
Afterwards the model is updated on the canvas.
Thus the modification of client A and the extensions of the event system could be transferred to client B.
At the same time, it becomes clear that conflicts can arise at various points, which will be explained in the next section.

\subsection{Concurrency Control}

The distributed system generated by Pyro allows users to make changes to the same model in parallel.
Through the underlying BASE communication model, the state is initially distributed to a client and then only changes encoded as CRDTs are transmitted to the server.
Although this concept achieves high availability and scalability, concurrent modifications are also possible, which must be handled as part of a concurrency control.
\begin{figure*}
	\centering
	\includegraphics[width=0.9\textwidth]{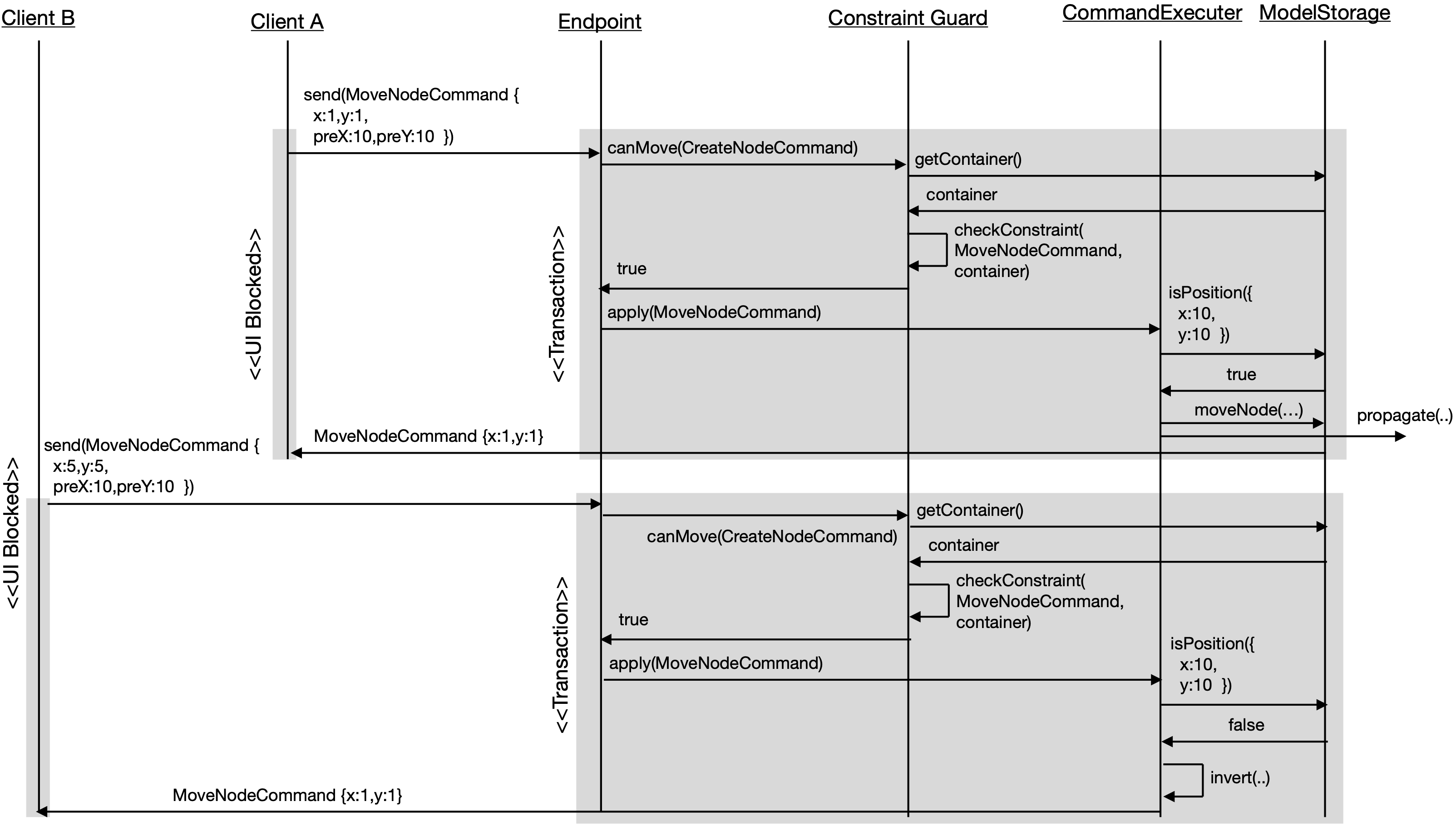}
	\caption{Concurrent client to server write conflict detection.}
	\label{fig:server_conflict}
\end{figure*}
The distributed system generated by Pyro corresponds due to its architecture to the mathematical actor model by Hewitt \cite{hewitt2010actor} for concurrent computation by fulfilling the following properties:
\begin{itemize}
\item Both client and server correspond to an actor as a computational entity.
\item Each actor sends and receives messages
\item Received messages can change the local state of an actor.
\item Actors can influence each other exclusively via messages.
\end{itemize}
The restriction to message passing for asynchronous communication between actors eliminates the need for lock-based synchronization \cite{hewitt2010actor} , which improves the local modification of a model in the client with respect to availability.
At the same time, the order in which messages are received and processed does not matter as long as the part of the state affected by the message has not been modified before.
To handle this readers-writers problem \cite{courtois1971concurrent} and to maintain consistency, an additional concurrency control must be implemented within the server.

The concurrency control on the server ensures that correct results are generated for concurrent operations and that this is done as quickly as possible \cite{kung1981optimistic}.
Conceptually, each client is a separate component that operates correctly on its own private state and maintains consistency.
However, since the components interact asynchronously and in parallel by exchanging messages with the server and thereby accessing and modifying shared data, it is possible that consistency is violated.
As described in the CAP theorem \cite{brewer2000towards}, when using a distributed data store, it is not possible to simultaneously provide more than two of the three guarantees of consistence, availability, and partition tolerance.
With respect to the server generated by Pyro, consistency is periodized so that:
\begin{itemize}
\item each read operation preserves the last modified state
\item and the system operates continuously even if operations may have to be rejected.
\end{itemize}
Thus, the concurrency control takes place exclusively on the server, rejecting the submitted operations in case of conflict to maintain consistency at the expense of availability.
The control mechanism here is transactional optimistic and non-blocking, in that an operation is executed until an integrity violation occurs or the transaction is terminated which has been described by Guerraoui \cite{guerraoui2002non}.
In the event of a violation, the transaction is fully rolled back in terms of an atomic commit and an error is returned.
With regard to the communication process described in section \ref{sec:communication_process}, various points can be identified which potentially endanger the consistency of the central model.

\subsubsection{Concurrent Server Write}

Fig. \ref{fig:server_conflict} shows an example of the readers-writers problem according to Courtois et al. \cite{courtois1971concurrent} which occurs when a client B tries to modify a shared resource after it has already been modified by another client A without client B having seen the modification before.
The given example starts with client A moving a node from position $(10,10)$ to $(1,1)$.
This operation is encoded by a corresponding message with the previous and new position of the node and sent to the server.
The server first checks whether the operation violates the syntactic contraints from the metamodel and then executes the operation.
From this point on, the centrally stored model is modified and the change is propagated to all listening clients.
However, it is possible that another client B has moved the same node to a different position before receiving this change.
As seen in the example, the transmitted message is first validated for syntactic correctness as before.
Then, an attempt is made to apply the operation on the central model.
Before this, however, it is checked whether the previous position passed in the message matches the centrally stored one.
If it does not, the transaction is aborted at this point and a revert message is sent back to the corresponding client.
Thus, Pyro implements the write-repair mechanism \cite{pritchett2008base} of the BASE communication model by making corrections to the client state in case of incorrect writes.

Other possible inconsistencies that may occur, for example, due to the ABA problem \cite{dechev2010understanding}, are not considered in the context of Pyro, since an operation is considered valid as long as the state on a client matches that of the server.
This means that an intermediate change that has been reverted does not imply an inconsistency.
Other dead-lock related concurrency problems like the dining-philosophers problem by Dijkstra \cite{dijkstra1971hierarchical} are completely avoided by not using locks. 

\subsubsection{Concurrent Client Write}

Due to the distribution of operations messages from the server to the listening clients, it can happen that a conflict exists for the local MirrorModel state.
This conflict occurs when an element has been processed by the user and an operation related to the same element is received from the server at the same time.
\begin{figure}
	\centering
	\includegraphics[width=0.45\textwidth]{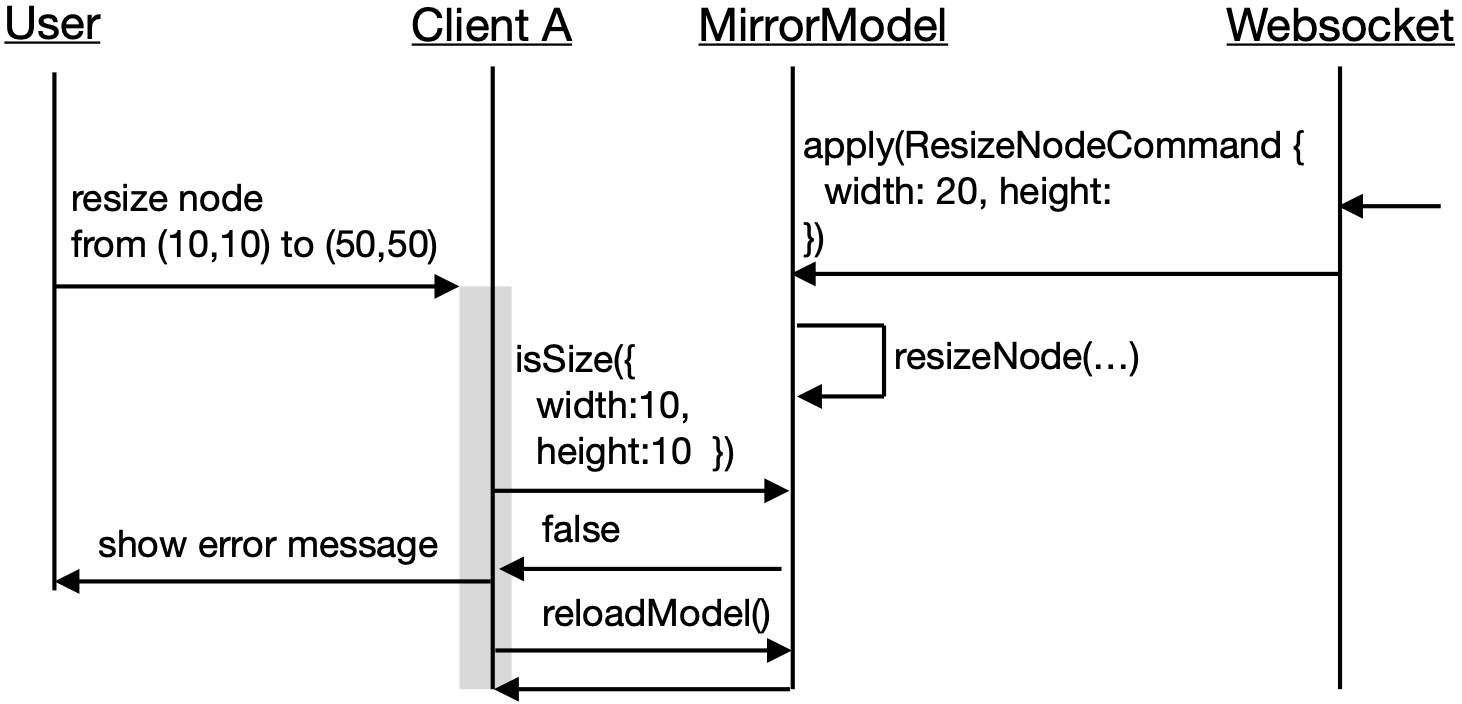}
	\caption{Concurrent server to client write conflict detection.}
	\label{fig:client_conflict}
\end{figure}
Fig. \ref{fig:client_conflict} visualizes such a conflict situation in an example.
A user changes the size of a node on his client via the canvas.
Before the operation can be applied by the on the local MirrorModel, a message arrives from the server that resizes the same node in a different way.
Thus, the state of the node in the MirrorModel is different from the one before the user made his change on the canvas.

However, this difference is checked as soon as the event system behind the canvas tries to apply an operation locally.
In case of a conflict, the operation is discarded and instead the canvas is synchronized with the current state of the MirrorModel that contains the change received from the server. 
Thus, the message from the server is always given priority and the user's local change is discarded by using the write-repair mechanism at this point as well.
This step is necessary to obtain the additional safety guarantee of the strong eventually consistency model \cite{vogels2009eventually} by ensuring that two clients have the same local state after receiving the same messages.

\section{Tool Usage}
\label{sec:user_interface}

The user interface of a modeling platform must meet various requirements to support a user not only in editing a model but also in managing it:
\begin{itemize}
\item Users must be able to flexibly manage organization and project structures.
\item It must be ensured that certain users have only certain influence possibilities.
\item The user interface must optimally support the user in editing the DSL. 
\end{itemize}
To meet these requirements, the user interface of the generated modeling environment offers different interfaces for organization, project and user management and the central editor.
\begin{figure*}
	\centering
	\includegraphics[width=0.9\textwidth]{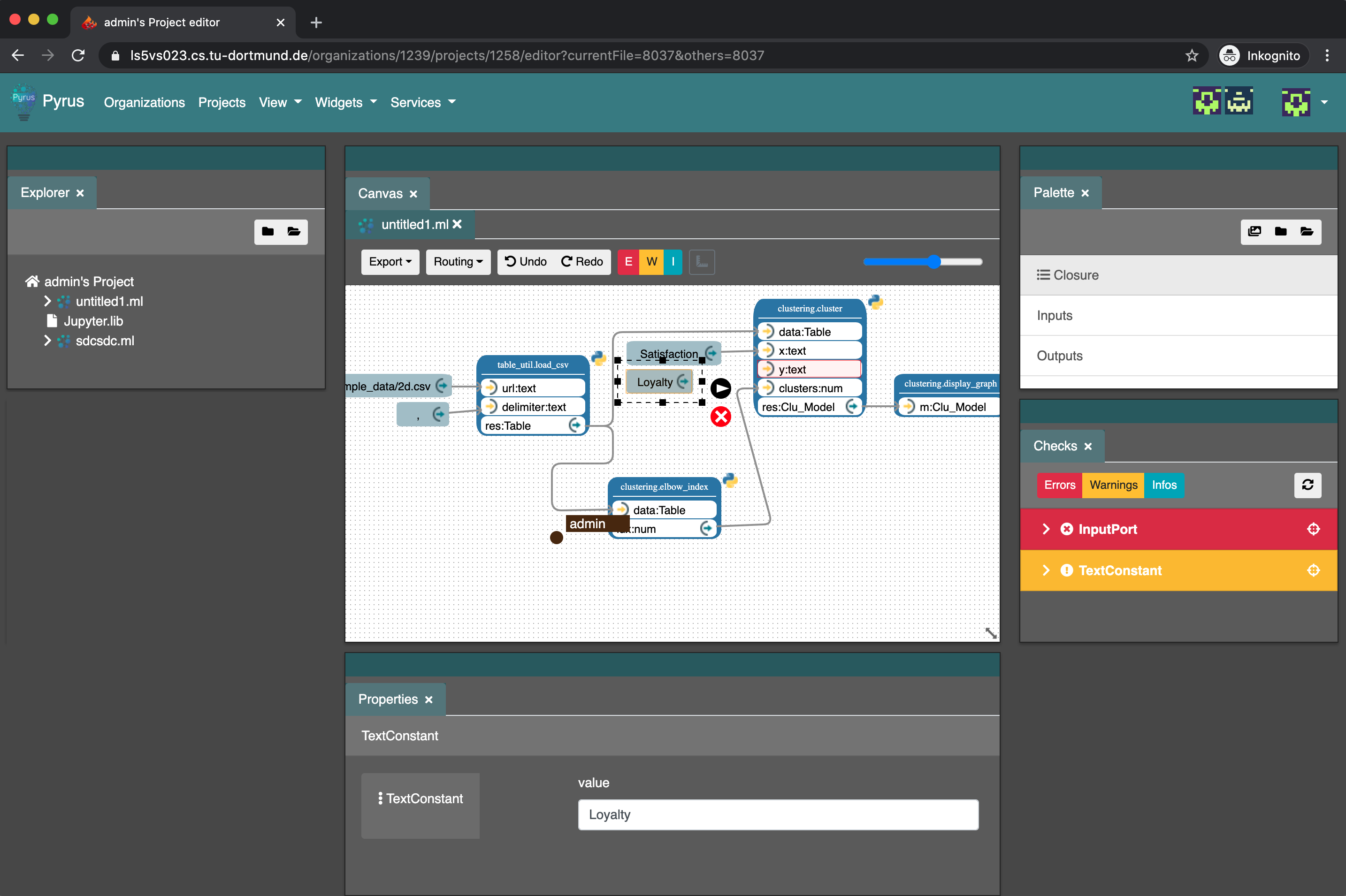}
	\caption{Generated Pyrus editor user interface.}
	\label{fig:screen}
\end{figure*}
\subsection{Management}

The management interfaces are used to allow users to register and log in.
For better structuring, users can divide themselves into organizations and manage their projects there.
In addition, Pyro offers an integrated roles and rights management in which per organization can be defined which users have which administrative and DSL related possibilities.

The possible rights refer to the basic create read update and delete (CRUD) operations a user can perform on the organization, on a project or on a model.
Initially, the owner who created the organization has all rights and can add other users who initially only have read rights.
Via the corresponding administration interfaces, so-called organization members can be given further rights.

Within an organization, projects can be created which are instantiated based on the previously specified UIProfiles (see section \ref{sec:user_interface_design}).
Depending on the previously made specification, the corresponding DSL types can be created and edited within the project.
Each project is to be understood as an independent virtual root folder, in whose hierarchy all resources of the project are stored.
Users within the organization can join and participate in a project according to their rights.

\subsection{Editor}

The project itself and all resources within it can be managed and edited via the editor.
The user interface of the editor is pre-initialized by the specified UIProfile.
Fig \ref{fig:screen} shows the example editor user interface for the Pyrus tool developed using Pyro.
The Pyrus tool provides a DSL for data-driven composition of externally specified functions.

As specified in the corresponding UIProfile, there is a navigation bar at the top to return to the previous management interfaces as well as the possibility to manage the current user account.
On the left below is the explorer, which displays the folder structure of the current project and allows the creation of additional models and resources.

In the center, the currently opened model is displayed on a canvas.
All nodes and edges of the model are visualized, which can be edited directly via drag'n'drop.
By clicking on an element, the corresponding node or edge is selected and a menu for editing is displayed.
As shown in Fig. \ref{fig:screen}, the menu of a node allows to drag edges using the black button and to resize the shape.
At the same time, a Properties View is displayed below the canvas for the selected element.
The Portperties View provides a form to assign values to the attributes specified in the metamodel of the DSL.
Above the canvas there are further model overlapping features to influence for example the edge layout or export the model as an image.

Besides the support of the individual user, the editor also offers additional visualizations for collaboration.
For example, the cursor positions of the users working on the model in parallel are displayed on the canvas, so that it is clear where changes are currently being made.
At the same time, the right end of the navigation bar shows which users are currently working on the project.
This non-functional information is continuously updated and improves the overview during simultaneous collaboration.  

To create additional nodes, the entries of the palette can be dragged from the right side onto the canvas.
The palette is created depending on the metamodel and lists all existing nodes and container types.
The entries can be grouped by the DSL metamodel to help the user find a specific type.

Below the palette is the Check View, which displays the results of the model validation.
The validation is performed in parallel with the modeling and displays the results of a syntactic and semantic checks.
In addition, the corresponding elements are also colored accordingly on the canvas, so that the user is guided directly to the problematic parts of the model.
The modeling of a DSL can thus be done completely via the editor, which uses the interaction design of drag'n'drop to enable intuitive editing for the user.

\section{Related Approaches}
\label{sec:tool_comparison}

Several frameworks and language workbenches exist for the creation and delivery of graphical domain-specific modeling languages.
However, these tools do not focus on the special requirements of non-technical users although they represent the ideal target group for domain-specific tools.
Thus, the comparable approaches address parts of the required features, but never the full scope.
In particular, the features of direct usability and simultaneous collaboration in combination with the simplicity-driven creation of new DSLs accordingly represent a unique selling point of Pyro.

In this section, the most widely used approaches are analyzed and compared with each other with respect to the current requirements of non-technical users described earlier in section \ref{sec:user_requirements}.
The analysis is limited to tools that provide direct usability and simultaneous collaboration.
The comparison has been limited to tools that provide exclusively web-based access to modeling and are restricted to graphical DSLs.
We provide our assessment and discussion based on previous experience with the tools as well as experiments conducted explicitly for this article.
A full analysis based on more general criteria and metrics, as presented in \cite{iung2020systematic,gray2016evolution}, is beyond the scope of this article.

In the next sections, Pyro is compared to the following tools and frameworks:
\begin{table*}
	\begin{tabularx}{\textwidth}{|l||X|X|X|X|}
		\hline
		Feature \quad \textbackslash \quad Framework & WebGME & Sirius Web & Eclipse Cloud Development &  Pyro \\ \hline
		\hline
		\ref{subsec:meta_layer_separation} Meta Layer Separation & no  & by configuration & by implementation & by generation \\ \hline
		\ref{subsec:ui_ixd_specialization} UI and IxD Specialization & no & no & no & yes \\ \hline
		\ref{subsec:too_developer_support} Tool Developer Support & generic & generic & generic & specialized \\ \hline
		\ref{subsec:simultaneous_collaboration} Simultaneous Collaboration & pessimistic, none-blocking & pessimistic, blocking & no & optimistic, none-blocking \\ \hline
	\end{tabularx}
	\caption{Comparison of web-based graphical domain-specific modeling tools and frameworks.} \label{tbl:comparison}
\end{table*}

\emph{WebGME} \cite{maroti2014next} is a versatile and generic tool for the simultaneous creation and use of domain-specific graphical languages on the web.
A user is thus able to edit both the metamodel of a DSL and instances of it.
The created models are stored and managed centrally on a server.
WebGME offers simultaneous collaboration during modeling and manages additional version histories to allow branching and snapshots.
Conflict handling is accordingly defensive, with each potential conflict creating a new branch.
WebGME's user interface provides each user with the full functionality of the tool.
Users can be managed and assigned to projects within the WebGME application.
In addition, WebGME provides special concepts for the creation of new DSL by defining prototypical inheritance and crosscuts that support meta-modeling and extend the approach of AToMPM \cite{syriani2013atompm}. 

\emph{Sirius Web} \cite{wwwsiriusweb} is the web-based extension of the Eclipse Sirius Project (Sirius Desktop) \cite{wwwsiriusdesktop} as part of the Eclipse Modeling Project.
Sirius Web uses a metamodel previously created with Sirius Desktop and generates an intermediate representation which is interpreted by Sirius Web to configure a generic web application.
The user interface and interaction design of a created DSL was modeled after the Eclipse-based interface by allowing different views to be used for a model.
In contrast to the established freehand model editors, Sirius was based on the so-called assisted layout which positions all elements in a previously defined grid.
Within the assisted layout Sirius Web supports the simultaneous collaboration between different users on one model.

\emph{Eclipse Cloud Development (ECD)} \cite{wwwecd} is a collection of projects, technologies and platforms to enable web-based software development in the cloud.
With regard to model-driven software development with graphical DSLs, the projects EMF cloud \cite{wwwemfclous}, GLSP \cite{wwwglsp}, Sprotty \cite{wwwsprotty} and Theia \cite{wwwtheia} are particularly relevant, which can be combined to realize a web-based modeling environment.
EMF cloud is used for central persistence of a model and Theia in connection with Sprotty as editor.
The link is the Eclipse graphical language server protocol, which describes an operation-based procedure how changes can be exchanged between server and client.
Based on the concept of the Language Server Protocol (LSP) \cite{bunder2019decoupling} the necessary mechanisms for the use of a DSL are centralized on a server.
By using the GLSP and Sprotty it is possible to choose between different generic editors like Theia and Eclipse to use a DSL.
In contrast to the other tools the projects of the ECD are not supported by a language workbench, so that the realization of a new DSL must take place for the most part manually.

Table \ref{tbl:comparison} summarizes our comparison in terms of the requirements of non-technical DSL users.
Each of the following sections explains one of the rows of the table.

\subsection{Meta Layer Separation}
\label{subsec:meta_layer_separation}

Domain-specific tools have the central task of optimally supporting the user in the respective domain.
In order to achieve this goal, only necessary features may be present in a tool, to not confuse or overwhelm the user.
Even if it can be advantageous for experienced users to change a DSL independently by adjusting the metamodel, this additional dimension of complexity is overwhelming and unnecessary for non-technical users.
Therefore, in our view, it does not make sense to manage the metamodel of a DSL at the same level as the DSL itself.

In this point the approaches of Pyro, Sirius Web and Eclipse Cloud Development agree in that the development of the language takes place on a separate meta-level to specify a tool.
However, only Pyro and Sirius Web have an approach where a tool can be created directly from a specification.
The ECD, on the other hand, requires that a DSL has to be implemented manually based on the framework.

The alternative approach of WebGME requires that the metamodel of a DSL must be managed on the same level as its instances.
This results in a much more complex tool that allows faster changes to a DSL but at the cost of simplicity and intuitiveness.

\subsection{UI and IxD Specialization}
\label{subsec:ui_ixd_specialization}

In addition to the DSL as the core of the tool, it is equally important for user acceptance that the surrounding user interface can be adapted to the needs of the users.
This can refer to the existing widgets and components as well as the layout and styling of the interfaces.
Similarly, the interaction design should also be tailored as much as possible by influencing the way a model is edited during tool development.

In this respect, Pyro's approach differs from both WebGME and Sirius Web.
WebGME is based on a generic user interface that is the same for all developed DSL.
Since WebGME does not create a specific tool per DSL, it is not possible to develop different user interfaces or interactions for one DSL or set of users.

Similarly, the editor used by Sirius Web cannot be customized for a specific user group.
The use of a generic interpreter, which reads out the previously specified metamodel at runtime, creates a user interface that is always the same.
Even if it would be possible in principle to extend the interpreter by a multiplicity of possible parameters for the user interface description it is not yet done at the present time.
In addition, Sirius Web only offers the possibility of the so-called assisted layout, so that it is not possible for a user to position nodes and edges independently and to work freehand.

The use of ECD projects makes it possible in principle to develop a completely specific user interface.
However, this must be done completely by hand and iteratively for each additional DSL.
However, under the premise of LDE and XMDD that a domain-specific tool should be as specialized as possible, it is necessary to find an efficient way to create such amount of required tools.

Pyro, on the other hand, addresses the specialization of the UI and IxD in an additional metmodel besides the DSL definition.
In addition to this, the user interface can be specified specifically for one or more user communities.
Likewise, in parallel to the language metamodel, mechanisms such as the event system can be extended to tailor the processing of a model to the domain.
The subsequent automatic generation step completely generates the corresponding specific tool by scaling the development in terms of XMDD.

\subsection{Tool Developer Support}
\label{subsec:too_developer_support}

When providing a framework, it is necessary to simplify the development as much as possible and at the same time provide sufficient specialization options.
This becomes particularly relevant when it is necessary to implement extensions for a domain-specific tool.
All presented approaches allow in principle to implement extensions for a DSL such as validators, generators, interpreters or an event system.
However, the presented tools differ in the level of support for a tool developer.

Sirius Web provide declarative support for the specification of extensions.
These can be specified at the meta-level in Eclipse via a configuration file.
However, the possible extensions are limited to previously specified locations.
As soon as additional extensions such as another context menu action are necessary, a developer would be forced to extend Sirius himself.

By managing metamodel and instances in parallel, WebGME is based on a generic core, which can be extended by existing extension points or an web API.
Even though arbitrary extensions can be developed in this way, developers are forced to operate on the generic API of WebGME without a reference to the actual DSL.

The ECD projects also offer arbitrary extension possibilities, however, also with this framework exists only a generic API and no support by a language workbench is given.

In contrast, Pyro follows the full generation approach, where the API of the tool is also generated based on the metamodel of the DSL.
Thus, Pyro allows extensions to be developed on top of a specialized API, which both eases the entry point and increases the efficiency of development.

\subsection{Simultaneous Collaboration}
\label{subsec:simultaneous_collaboration}

One of the main advantages of web-based IDEs for a user is the possibility to work simultaneously on one model.
Therefore, collaboration is a central requirement for a tool, which must be offered independently of the targeted domain.
At the same time, it is also necessary that synchronization does not create additional effort for the user or interfere with his way of working.
Measured against established non-technical tools outside MDSD, such as Google Docs \cite{wwwgoogledocs}, users expect synchronization to work without conflicts or conflict handling as far as possible.

Contrary to this requirement, WebGME implements a pessimistic collaboration mechanism in which a branch is created in the history for each potential conflict.
Even if this system avoids conflicts in general and does not block the user, it is necessary to merge versions afterwards within a time-consuming merge process.

Sirius Web implements a thin-client approach, so that all operations must be validated by the server before they become visible to a user.
However, in conjunction with the assisted layout mechanism, this results in a pessimistic workflow characterized by interruptions during modeling, in that the user must wait until a response is received from the server after each interaction with the tool.
In case of a conflict, the user must directly perform a merge of the divergent versions. 

Since the development of the Eclipse GLSP is still in an early stage of development and the initial focus has been on broad usage and integration with different editors, simultaneous collaboration is not yet supported.

A tool generated by Pyro is based on the BASE communication model, which acts optimistically by making all actions visible directly on a client's canvas and only checking global consistency in a subsequent step.
Although this approach may result in the need to revert a user's action in the event of a conflict, Pyro provides a transparent and non-disruptive way of working for the user.

\section{Conclusion}
\label{sec:summary_and_outlook}

We presented Pyro, which can be used to create holistic web-based domain-specific modeling tools.
For this purpose Pyro implements approaches like the one-thing approach and full code generation by starting from different declarative specifications and generating a complete modeling environment.
The key to the Pyro idea is the meta-level based tool development through the metamodel of the DSL and UI as well as additional extensions like the interpreter framework to breathe life into a language.

In contrast to classical integrated development environments, Pyro focuses on the target group of non-technical users.
For this purpose, we first analyzed in detail which special requirements are placed on such a tool and which technical challenges arise from this.
The target user group expects graphical, directly usable, specific and collaborative tools, which Pyro offers through the generative realization of a distributed server-client modeling environment.

The generated system will be completely specialized for a domain, starting with the persistence of a model in the database, over the controller layer, up to the editor of the client.
A special unique selling proposition is represented by the used communication model, which is optimistically designed in contrast to the other existing tools.
By implementing BASE and avoiding locks, a user gets the feeling of unrestricted simultaneous collaboration without the need to deal with version control systems.

Overall, we are constantly evolving Pyro and its underlying concepts.
The generative approach allows these changes to take place exclusively at the meta-level, and in the next step to be mostly automatic for all tools created up to that point.

In view of the growing use of web-based domain-specific tools, we try to combine the generative approach of Pyro with the Eclipse Cloud Development projects.
Since the ECD is mainly focused on classical generic software development, we hope to enable the best of both worlds by combining both technologies.
For example, by uniting the Eclipse GLSP and the Pyro GLSP, the collaborative usability of a DSL into different editors could be achieved.

\bibliographystyle{spmpsci}      

\bibliography{references}

\end{document}